\documentclass{article}

\usepackage{hyperref}

\usepackage[accepted]{icml2026}

\usepackage{minted}

\usepackage{xspace}
\usepackage[utf8]{inputenc} 
\usepackage[T1]{fontenc}    
\usepackage{hyperref}       
\usepackage{url}            
\usepackage{booktabs}       
\usepackage{amsfonts}       
\usepackage{nicefrac}       
\usepackage{microtype}      
\usepackage{xcolor}         
\usepackage[table]{xcolor}
\usepackage{soul}
\usepackage{enumitem}
\usepackage[most]{tcolorbox}
\usepackage[tikz]{bclogo}
\usepackage{wrapfig}
\usepackage{multirow}
\usepackage{multicol}
\usepackage{hhline}
\usepackage{tabularray}
\usepackage{booktabs}
\usepackage{arydshln}
\usepackage{tablefootnote}
\usepackage{pifont}
\usepackage{changepage}

\usepackage{microtype}
\usepackage{graphicx}
\usepackage{subcaption}
\usepackage{booktabs} 

\usepackage{amsmath}
\usepackage{amssymb}
\usepackage{mathtools}
\usepackage{amsthm}

\usepackage[capitalize,noabbrev]{cleveref}

\theoremstyle{plain}

\theoremstyle{definition}

\theoremstyle{remark}

\usepackage[textsize=tiny]{todonotes}

\tcbuselibrary{listingsutf8}

\newcommand{\approach}{\textsc{Match\-Fix\-Agent}\xspace}
\newcommand{\alphatrans}{\textsc{AlphaTrans}\xspace}
\newcommand{\oxidizer}{\textsc{Oxidizer}\xspace}
\newcommand{\rustrepotrans}{\textsc{RustRepoTrans}\xspace}
\newcommand{\skel}{\textsc{Skel}\xspace}
\newcommand{\syzygy}{\textsc{Syzygy}\xspace}

\newcommand*\circled[2][fill=black]{\tikz[baseline=(char.base)]{\footnotesize \node[shape=circle, #1, inner sep=1pt] (char) {\textcolor{white}{#2}};}}

\newtcolorbox{monotextbox}{
    colback=yellow!10,
    colframe=black,
    boxrule=0.5pt,
    arc=2pt,
    left=4pt,
    right=4pt,
    top=2pt,
    bottom=2pt,
    fontupper=\ttfamily\footnotesize,
    enhanced,
}

\definecolor{problemblue}{RGB}{100,134,158}
\definecolor{idiomsgreen}{RGB}{0,162,0}
\definecolor{exercisebgblue}{rgb}{0,  .69,  .941}
\definecolor{deepgreen}{rgb}{0.0, 0.5, 0.0}
\definecolor{codegreen}{rgb}{0,0.6,0}
\definecolor{codegray}{rgb}{0.5,0.5,0.5}
\definecolor{codepurple}{rgb}{0.58,0,0.82}
\definecolor{backcolour}{rgb}{0.95,0.95,0.92}
\definecolor{redColor}{RGB}{255,0,0}
\definecolor{Gray}{gray}{0.1}
\definecolor{bubblegum}{rgb}{0.99, 0.76, 0.8}
\definecolor{cambridgeblue}{rgb}{0.64, 0.76, 0.68}
\definecolor{babypink}{rgb}{1.0, 0.82, 0.86}
\definecolor{lightcoral}{rgb}{0.94, 0.66, 0.66}
\definecolor{mistyrose}{rgb}{1.0, 0.89, 0.88}
\definecolor{orchidpink}{rgb}{0.95, 0.74, 0.80}
\definecolor{carnationpink}{rgb}{1.0, 0.65, 0.79}
\definecolor{ashgray}{rgb}{0.70, 0.75, 0.71}
\definecolor{celadon}{rgb}{0.67, 0.88, 0.69}
\definecolor{powderblue}{rgb}{0.69, 0.88, 0.90}
\definecolor{etonblue}{rgb}{0.59, 0.78, 0.64}
\definecolor{teagreen}{rgb}{0.82, 0.94, 0.75}

\newcommand{\killpunct}[1]{}

\begin{document}

\twocolumn[
  \icmltitle{\approach: Language-Agnostic Autonomous Repository-Level Code Translation Validation and Repair}

  \icmlsetsymbol{intern}{$\dagger$}

  \begin{icmlauthorlist}
    \icmlauthor{Ali Reza Ibrahimzada}{uiuc,intern}
    \icmlauthor{Brandon Paulsen}{amazon}
    \icmlauthor{Reyhaneh Jabbarvand}{uiuc}
    \icmlauthor{Joey Dodds}{amazon}
    \icmlauthor{Daniel Kroening}{amazon,oxford}
  \end{icmlauthorlist}

  \icmlaffiliation{uiuc}{Siebel School of Computing and Data Science, University of Illinois Urbana-Champaign, Urbana, IL, USA}
  \icmlaffiliation{amazon}{Amazon, Arlington, VA, USA}
  \icmlaffiliation{oxford}{University of Oxford, Oxford, UK}

  \icmlcorrespondingauthor{Ali Reza Ibrahimzada}{alirezai@illinois.edu}

  \icmlkeywords{Program Analysis, Neuro-Symbolic Code Translation Validation and Repair, LLM Agent}

  \vskip 0.3in
]

\printAffiliationsAndNotice{$^\dagger$Work done when author was an intern at AWS.}

\begin{abstract}
  Code translation transforms source code from one programming language (PL) to another. Validating the functional equivalence of translation and repairing, if necessary, are critical steps in code translation. Existing automated validation and repair approaches struggle to generalize to many PLs due to high engineering overhead, and they rely on existing and often inadequate test suites, which results in false claims of equivalence and ineffective translation repair. To bridge this gap, we develop \approach, a large language model (LLM)-based, PL-agnostic framework for equivalence validation and repair of translations. \approach features a multi-agent architecture that divides equivalence validation into several sub-tasks to ensure thorough and consistent semantic analysis of the translation. We compare \approach's validation and repair results with four repository-level code translation techniques. Our results demonstrate that \approach produces (in)equivalence verdicts for $99.2\%$ of translation pairs, with the same equivalence validation result as prior work on $72.8\%$ of them. When \approach's result disagrees with prior work, we find that $60.7\%$ of the time \approach's result is actually correct. In addition, we show that \approach can repair $50.6\%$ of inequivalent translation, compared to prior work’s $18.5\%$. 

\end{abstract}

\vspace{-20pt}

\section{Introduction}
\label{sec:introduction}

\begin{figure*}[t]
    \centering
    \includegraphics[width=0.9\textwidth]{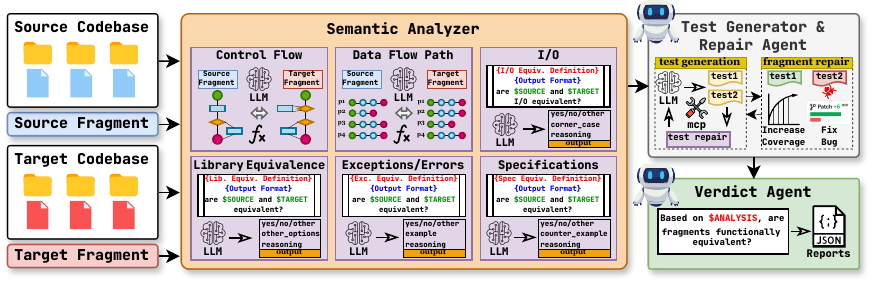}
    \vspace{-5pt}
    \caption{Overview of \approach.}
    \label{fig:overview}
    \vspace{-15pt}
\end{figure*}

Code translation, the process of converting source code from one programming language (PL) to another, is a cornerstone of software modernization efforts that enhance performance, maintainability, and reliability~\cite{khan2022modernization,jain2015modernization,jamshidi2013cloud,khadka2014professionals}. Translation validation and repair are integral steps in code translation for determining functional equivalence and patch generation for incorrect translations. However, performing validation and repair manually---particularly in large codebases---can be tedious, time-consuming, and error-prone, especially when complex code structures and dependencies are involved~\cite{hora2015developers,kula2018developers,wang2020empirical}. Prior work
defines value and type equivalences and translations to compare source and target implementations over pairs of concrete inputs. The inputs either come from existing tests from the source project~\cite{shetty2024syzygy, zhang2025oxidizer, wang2025skel, ibrahimzada2025alphatrans} or differential fuzzing~\cite{yang2025vert, eniser2024towards}. Despite notable advancements in validation and repair of repository-level code translation, existing techniques are hampered by the following limitations (examples given in Appendix~\ref{sec:limitations}): 

\begin{enumerate}[leftmargin=*]

    \vspace{-10pt}

    \item \emph{Difficulty Generalizing to Many PL Pairs.} While the fundamental ideas of current validation approaches extend to many PL pairs, their actual implementations typically support just one language pair. This is because supporting language interoperability between a pair of PLs requires a large engineering effort, as evidenced by the size of these tools\footnote{The implementation of notable recent translation and validation techniques are \alphatrans ($10859$ LoC), \oxidizer ($19052$ LoC), and \skel ($3843$ LoC).}.
    Given the quadratic number of PL pairs, language interoperability techniques are extremely challenging to scale to many PL pairs. 

    \vspace{-5pt}

    \item \emph{Unknown Test Requirements.} Current translation validation approaches require a set of valid inputs to validate the input-output equivalence between source functions and their translation. They generate these inputs by either executing available source tests~\cite{wang2025skel, zhang2025oxidizer, ibrahimzada2025alphatrans, shetty2024syzygy} or fuzzing the source project~\cite{eniser2024towards, yang2025vert} 
    Unit tests are often incomplete, missing important inputs, and resulting in false claims of equivalence. Fuzzing techniques suffer from generating invalid inputs, also resulting in false claims of inequivalence, and in general, failing to reach deep into the code or create complex objects in the context of real-world projects~\cite{klees2018evaluating,godefroid2008automated}. 

    \item \emph{Ineffective Translation Repair.} Recent studies have shown that a more rigorous validation can reveal more translation bugs~\cite{pan2024lost}.
    Hence, advancement in translation validation should be accompanied by effective repair strategies. Existing techniques, however, either require the user to fix incorrect translations manually~\cite{wang2025skel} or use feedback-driven re-prompting strategies~\cite{zhang2025oxidizer,ibrahimzada2025alphatrans} that are barely effective in repository-level code translation due to long call-chain dependencies in real-world projects~\cite{ibrahimzada2025alphatrans}.

\end{enumerate}


\vspace{-10pt}
Prior work consists of high-effort implementations that can deliver inconsistent results and only apply to one of a quadratic number of language pairs. Large Language Models (LLMs) have recently been successful at same-language equivalence validation~\cite{wei2025equibench,maveli2025can}, so replacing cross-language equivalence implementations with LLM decisions is a logical next step. While there is a risk of incorrect results, the baseline mechanical approaches already display low accuracy. 

Towards this end, we propose \approach, a \emph{language-agnostic} approach to automate \emph{validation} and \emph{repair} of repository-level code translation (\S \ref{sec:overview}). \approach combines the generative power of LLMs with several approximate code semantic analyses, i.e., analysis of control- and data-flow paths, library APIs, exception handling, and (formal or informal) specification (\S \ref{subsec:semantic-analyzer}). 
These semantic analyses are then fed to a \emph{test generator \& repair agent} to generate tests for assessing functional equivalence, and repair the translation in the case of failing tests (\S \ref{subsec:test-gen-repair}). 
The final equivalence decision will be made by the \emph{verdict} agent, considering the approximate semantic analyses and test execution results (\S \ref{subsec:verdict}). \approach is very lightweight ($1{,}650$ lines of code), modular, and interoperable with existing repository-level translation systems. 

We evaluate the effectiveness of \approach for repository-level translation validation and repair against four existing techniques~\cite{ibrahimzada2025alphatrans,zhang2025oxidizer,wang2025skel,ou2024repository}. Our benchmark comprises $2{,}219$ source–translation function pairs, which cover $6$ PL pairs, drawn from $24$ real-world projects totaling over $900K$ lines of code (details in Appendix \ref{appendix:experiment-setup}). For each translation pair, we obtain an equivalence verdict (validation outcome) from both \approach and other techniques. Overall, \approach returns a verdict for $99.2\%$ of pairs, while alternative approaches do so for only $71.6\%$ (\S \ref{subsec:rq1}). On the $1{,}571$ pairs where both produce verdicts, \approach agrees with other approaches in $72.8\%$ of cases. For the remaining disagreements, a systematic manual investigation finds \approach to be correct in $60.7\%$ of cases and incorrect in the rest (\S \ref{subsec:rq1-dispute-analysis}). In translation repair, \approach can fix $50.6\%$ of translation bugs, $32.1\%$ more than existing approaches (\S\ref{subsec:rq2}).
We show that \approach is compatible with different LLMs and agent frameworks, producing comparable results (\S \ref{subsec:rq3}). Lastly, our ablation study shows that removing code analyses and in-the-loop test generation reduces verdict accuracy by $42.3\%$, while increasing token usage by $5.2\%$ (\S \ref{subsec:rq4}). These results confirm that \approach is a viable alternative to prior work's validation and repair approaches, while being vastly easier to adapt to new PL pairs. Our notable contributions are:

\begin{enumerate}[leftmargin=*]
    
    \vspace{-10pt}
    \item We present \approach, a PL-agnostic, agentic approach for validation and repair of repository-level code translation.

    \vspace{-5pt}
    \item We demonstrate that \approach is a viable alternative to prior work's validation approach, while being vastly easier to adapt to new PL pairs.

    \vspace{-5pt}
    \item We show the benefit of \approach's multi-agent architecture compared to standalone-agent design.
    \vspace{-5pt}
\end{enumerate}

\section{\approach}
\label{sec:approach}

\subsection{Overview}
\label{sec:overview}

Figure~\ref{fig:overview} gives an overview of \approach, which consists of three main components: (1) the Semantic Analyzer~(\S\ref{subsec:semantic-analyzer}), (2) the Test Generator \& Repair Agent~(\S\ref{subsec:test-gen-repair}), and (3) the Verdict Agent~(\S\ref{subsec:verdict}). \approach takes as input a translation pair (source function and its translation) along with both projects, and outputs an equivalence verdict, a natural language report, and an optional repair patch. Algorithm~\ref{alg:approach} details the procedure.

The Semantic Analyzer invokes an LLM to analyze six semantic properties of the translation pair in parallel: control flow~(\S\ref{subsubsec:control-flow}), data flow~(\S\ref{subsubsec:data-flow}), input-output mapping~(\S\ref{subsubsec:io}), library API usage~(\S\ref{subsubsec:library}), exception handling~(\S\ref{subsubsec:exception-error}), and specifications~(\S\ref{subsubsec:spec}). This decomposition keeps the LLM focused and improves reliability~\cite{buildingeffectiveaiagents}. Each sub-task outputs a report summarizing semantic differences. These reports are fed to the Test Generator \& Repair Agent, which uses an off-the-shelf LLM coding agent (e.g., Claude Code~\cite{claudecode} or Codex~\cite{codex}) to write and execute tests that may reveal inequivalent behavior. If inequivalence is found, the agent attempts to repair the translation. Finally, the Verdict Agent validates the claims from the previous components and produces a final equivalence verdict with a summary.

\begin{algorithm}[t]
    \scriptsize

\begin{algorithmic}[1]
    \STATE {\bfseries Input:} $\mathit{sourceProject, sourceFunc, translatedProject,}$
    \STATE \hspace{3.2em} $\mathit{translatedFunc, LLM, tools, timeout}$
    \STATE {\bfseries Output:} $\mathit{validationRepairReport}$

    \STATE $\mathit{transPair} \leftarrow [\mathit{sourceFunc}, \mathit{translatedFunc}]$
    \STATE {\bfseries function} \textsc{semAnalyzer}($\mathit{transPair, LLM}$):
    \begin{ALC@g}
        \STATE $\mathit{cfgSrc, dfSrc} \leftarrow \textbf{build\_cfg}(\mathit{sourceFunc})$
        \STATE $\mathit{cfgTgt, dfTgt} \leftarrow \textbf{build\_cfg}(\mathit{translatedFunc})$
        \STATE {\bfseries return} \{
        \STATE \quad $\textbf{controlFlowAnalyzer}(\mathit{cfgSrc, cfgTgt, transPair, LLM}),$
        \STATE \quad $\textbf{dataFlowPathAnalyzer}(\mathit{dfSrc, dfTgt, transPair, LLM}),$
        \STATE \quad $\textbf{ioAnalyzer}(\mathit{transPair, LLM}),$
        \STATE \quad $\textbf{libraryAnalyzer}(\mathit{transPair, LLM}),$
        \STATE \quad $\textbf{exceptionAnalyzer}(\mathit{transPair, LLM}),$
        \STATE \quad $\textbf{specAnalyzer}(\mathit{transPair, LLM}) \}$
    \end{ALC@g}
    \STATE {\bfseries end function}
    \STATE \textbf{await} $\mathit{semAnalysis} \leftarrow \textsc{semAnalyzer}(\mathit{cfgSrc, dfSrc, cfgTgt, dfTgt})$
    \STATE $\mathit{testRepair} \leftarrow \textbf{testGenRepairAgent}(\mathit{prompt, LLM, tools, timeout})$
    \STATE $\mathit{verdict} \leftarrow \textbf{verdictAgent}(\mathit{semAnalysis, testRepair, LLM})$
    \STATE $\mathit{validationRepairReport} \leftarrow \mathit{semAnalysis} \cup \mathit{testRepair} \cup \mathit{verdict}$
    \STATE \textbf{return} $\mathit{validationRepairReport}$
\end{algorithmic}

    \caption{\small \approach}
    \label{alg:approach}
\end{algorithm}



\subsection{Semantic Analyzer}
\label{subsec:semantic-analyzer}

The Semantic Analyzer takes as input the translation pair and an LLM. It first computes a control flow graph (CFG) and data flow graph (DFG) (described in \S\ref{subsubsec:control-flow} and \S\ref{subsubsec:data-flow} respectively), and then calls six sub-analyzers in parallel, each of which analyzes a different semantic property of the translation pair. Each sub-analyzer invokes the LLM with a custom prompt
\footnote{Please refer to our artifacts repository~\cite{website} for prompt structure of each semantic analyzer.} 
describing the analysis to perform. The prompts are relatively simple and short. The prompt first defines a role for the LLM (\texttt{\small ``You are an expert in...''}), a general definition of functional equivalence, and a specific definition of equivalence for the semantic property. It then instructs the analyzer to output an equivalence verdict and explanation for the specific semantic property it analyzed. In addition, certain analyzers output examples to demonstrate inequivalence. The final output of the Semantic Analyzer is a 6-tuple containing a JSON-formatted output of each sub-analyzer. The following subsections provide more details on the prompts of the six sub-analyzers.

\subsubsection{Control Flow Analyzer}
\label{subsubsec:control-flow}

The \textit{Control Flow Analyzer} is prompted to analyze the control flow structures of the source and translation to determine whether they are equivalent, looking for inequivalences such as reordered conditions, missing branches, or altered loop termination criteria. 
To aid this task, we provide the LLM with textual representations of the source and translation's CFGs. An example is shown in Figure~\ref{fig:cfg-dfg-structure}. 
To compute the CFG, we use Tree-Sitter~\cite{tree-sitter} to construct an abstract syntax tree of the function, and then extract basic blocks and control flow structures. Tree-sitter supports $165$+ PLs, making this process PL-agnostic. Each PL supported by \approach required approximately $280$ lines of code, making it very easy to support many PLs. 

To improve reliability and reduce costs, the control flow analyzer first (symbolically) computes a similarity score between the CFGs, and, if it falls above a threshold, it immediately returns an equivalent verdict without invoking the LLM. The overall procedure is shown in Algorithm~\ref{alg:cfganalyzer}. The analyzer abstracts each graph into canonical forms (lines $3$--$13$) capturing node types (e.g., conditionals, loops, exception handlers) and edge types (control transfer relationships), then computes the structural similarity score based on the Jaccard index~\cite{cha2007comprehensive} (lines $14$--$16$). We use $0.7$ as the threshold. This results in approximately $25\%$ of LLM invocations being skipped in our experiments, making this threshold relatively stringent.

\subsubsection{Data Flow Analyzer}
\label{subsubsec:data-flow}

The \textit{Data Flow Analyzer} is prompted to evaluate whether the flow of data within the source and translation is equivalent, looking for issues like unused variables. 
To aid the LLM, we provide textual representations of the source and translation's data flow graphs (DFGs). An example is shown in Figure~\ref{fig:cfg-dfg-structure}.
We keep our data flow computation extremely simple. For each statement in the AST, we extract variable names, label them as a def or a use, and associate them with a CFG node. We then extract def-use chains. This is primarily a syntactic analysis. We do not handle challenging problems such as aliasing, concurrency, or context sensitivity. The per-PL implementation effort is approximately $280$ lines of code based on the six PLs supported by \approach. Similar to our control flow analyzer, we compute a similarity score between the DFGs, and short-circuit if it falls above a threshold. The analyzer, show in Algorithm~\ref{alg:dfpathanalyzer}, first extracts \emph{def-use} chains for parameters and local variables, capturing how data values are defined, propagated, and consumed. The extracted paths are compared using edit distance~\cite{miller2009levenshtein} as the similarity measure (lines $3$--$18$). We again use $0.7$ as the threshold, which results in approximately $35\%$ of LLM invocations being skipped.

\begin{algorithm}[t]
    \caption{\small Control Flow Analyzer}
    \label{alg:cfganalyzer}
    \scriptsize

\begin{algorithmic}[1]
    \STATE {\bfseries Input:} $\mathit{cfgSource, cfgTarget, fragments, model, \tau{=}0.7}$
    \STATE {\bfseries Output:} $\mathit{cfgAnalysis}$
    \STATE {\bfseries function} \textsc{abstractGraph}($\mathit{cfg}$):
    \begin{ALC@g}
        \FOR{\textbf{each} $(u, v)$ \textbf{with} $\mathit{edge} \in \mathit{cfg}$}
            \STATE $\mathit{uType, vType} \leftarrow \textbf{classifyNode}(u), \textbf{classifyNode}(v)$
            \STATE $\mathit{edgeType} \leftarrow \textbf{classifyEdge}(\mathit{edge})$
            \STATE $\mathit{nodes} \leftarrow \mathit{nodes} \cup \mathit{uType} \cup \mathit{vType}$
            \STATE $\mathit{edges} \leftarrow \mathit{edges} \cup \langle \mathit{uType, edgeType, vType} \rangle$
        \ENDFOR
        \STATE {\bfseries return} $\mathit{nodes, edges}$
    \end{ALC@g}
    \STATE {\bfseries end function}
    \STATE $\mathit{sourceNodes, sourceEdges} \leftarrow \textsc{abstractGraph}(\mathit{cfgSource})$
    \STATE $\mathit{targetNodes, targetEdges} \leftarrow \textsc{abstractGraph}(\mathit{cfgTarget})$
    \STATE $\mathit{nodeSim} \leftarrow \textbf{jaccardSimilarity}(\mathit{sourceNodes, targetNodes})$
    \STATE $\mathit{edgeSim} \leftarrow \textbf{jaccardSimilarity}(\mathit{sourceEdges, targetEdges})$
    \STATE $\mathit{similarity} \leftarrow (0.5 \times \mathit{nodeSim}) + (0.5 \times \mathit{edgeSim})$
    \IF{$\mathit{similarity} \ge \tau$}
        \STATE $\mathit{cfgAnalysis} \leftarrow \{ \text{"is\_equivalent": "yes"} \}$
    \ELSE
        \STATE $\mathit{cfgAnalysis} \leftarrow \textbf{LLM}(\mathit{cfgSource, cfgTarget, fragments, model})$
    \ENDIF
    \STATE {\bfseries return} $\mathit{cfgAnalysis}$
\end{algorithmic}

\end{algorithm}

\begin{algorithm}[t]
    \caption{\small Data Flow Path Analyzer}
    \label{alg:dfpathanalyzer}
    \scriptsize

\begin{algorithmic}[1]
    \STATE {\bfseries Input:} $\mathit{dfSource, dfTarget, fragments, model, \tau{=}0.7}$
    \STATE {\bfseries Output:} $\mathit{dfAnalysis}$
    \STATE {\bfseries function} \textsc{computeEditDistance}($\mathit{srcPath, tgtPath}$):
    \begin{ALC@g}
        \STATE $\mathit{sim\_a} \leftarrow 0$
        \FOR{\textbf{each} $\mathit{xPath} \in \mathit{srcPath}$}
            \STATE $\mathit{best} \leftarrow 0$
            \FOR{\textbf{each} $\mathit{yPath} \in \mathit{tgtPath}$}
                \STATE $\mathit{score} \leftarrow \textbf{jaccardSimilarity}(\mathit{xPath, yPath})$
                \STATE $\mathit{best} \leftarrow \textbf{max}(\mathit{best, score})$
            \ENDFOR
            \STATE $\mathit{sim\_a} \leftarrow \mathit{sim\_a} + \mathit{best}$
        \ENDFOR
        \STATE $\mathit{sim\_a} \leftarrow \mathit{sim\_a} / |\mathit{srcPath}|$
        \STATE $\mathit{sim\_b} \leftarrow$ \textit{repeat loop with srcPath and tgtPath swapped}
        \STATE {\bfseries return} $(\mathit{sim\_a} + \mathit{sim\_b}) / 2$
    \end{ALC@g}
    \STATE {\bfseries end function}
    \STATE $\mathit{srcPath, tgtPath} \leftarrow \textbf{getVariablePaths}(\mathit{dfSource, dfTarget})$
    \STATE $\mathit{similarity} \leftarrow \textsc{computeEditDistance}(\mathit{srcPath, tgtPath})$
    \IF{$\mathit{similarity} \ge \tau$}
        \STATE $\mathit{dfAnalysis} \leftarrow \{ \text{"is\_equivalent": "yes"} \}$
    \ELSE
        \STATE $\mathit{dfAnalysis} \leftarrow \textbf{LLM}(\mathit{dfSource, dfTarget, fragments, model})$
    \ENDIF
    \STATE {\bfseries return} $\mathit{dfAnalysis}$
\end{algorithmic}

\end{algorithm}

\subsubsection{IO Analysis}
\label{subsubsec:io}
The \textit{IO Analyzer} is prompted to evaluate whether the observable input-output behavior of the source and target fragments is semantically equivalent. 
The prompt includes an IO equivalence definition, which assess five dimensions: (1) semantic equivalence of accepted inputs, (2) consistency of produced outputs, (3) preservation of side effects (e.g., file operations, network calls, or global state modifications), (4) uniform handling of edge cases, and (5) similarity in performance-critical complexity. 
The LLM is prompted also prompted to produce a plausible input that would trigger dissimilar IO behavior if it believes the translation is inequivalent. 
This methodology catches inequivalences such as differing error messages, inconsistent encoding assumptions, or missing side effects---often overlooked by structural analyses (\S\ref{subsubsec:control-flow}, \S\ref{subsubsec:data-flow}) alone.

\subsubsection{Library Analyzer}
\label{subsubsec:library}
The \textit{Library API Analyzer} is prompted to consider the behavior of external library APIs called in the source and translation, and evaluate whether their differences result in inequivalent behavior. This analyzer primarily detects subtle differences between similar library APIs in the source and translation. It provides suggestions to fix inequivalent behavior as well.

\subsubsection{Exception \& Error Analyzer}
\label{subsubsec:exception-error}

The \textit{Exception and Error Handling Analyzer} is prompted to validate whether error detection, exception raising, and error recovery mechanisms in the source and target code fragments are functionally equivalent. The prompt include five dimensions for equivalence: (1) detecting and handling the same error conditions, (2) using semantically equivalent exception/error types, (3) producing equivalent error messages or codes, (4) preserving consistent recovery mechanisms, and (5) propagating errors in equivalent ways. If neither fragment implements explicit error handling, the analyzer deems them equivalent for this dimension. Otherwise, it statically identifies exception constructs (e.g., \texttt{\small try-catch}, \texttt{\small throw}, \texttt{\small return}-error patterns) and uses LLM reasoning to compare semantics. For instance, if the source raises a specific \texttt{\small FileNotFoundError} while the target raises a generic \texttt{\small IOException}, the discrepancy is flagged, as it may affect upstream handling. Where differences exist, the analyzer also recommends target-language error handling constructs that align with the source's semantics.

\subsubsection{Specifications Analyzer}
\label{subsubsec:spec}

The \textit{Specification Analyzer} is prompted to assess whether the source and target code fragments adhere to the same explicit or implicit functional specifications. 
The prompt includes Specification equivalence definition which state that the source and translation should: (1) fulfill the same documented or inferred functional requirements, (2) satisfy identical pre-conditions and post-conditions, (3) maintain the same invariants, and (4) handle the same input domain, including edge cases.
The LLM is instructed to extract available specifications from function signatures, type annotations, and relevant comments, or,
when no formal documentation exists, infer behavioral contracts from code semantics. 
The LLM is asked to compare the contracts of the source and translation. 
For example, if the source specifies \texttt{\small ``returns 1 on success, 0 on failure''} and the target returns \texttt{\small Boolean} values, the inconsistency is flagged. 
In such cases, the LLM is instructed to produce a formalized specification that reconciles both implementations and provides counterexamples demonstrating behavior mismatch.

\subsection{Test Generator \& Repair Agent}
\label{subsec:test-gen-repair}

The \textit{Test Generator \& Repair Agent} uses an off-the-shelf LLM coding agent and the reports from the Semantic Analyzer to write and execute tests that demonstrate functional (in)equivalence. This agent helps catch hallucinations and confirm the claims in the Semantic Analyzer reports. The prompt for the agent is shown in Figure~\ref{fig:test-gen-prompt-template}, which includes a definition of equivalence, instructions to write tests in both the source and target PL that test the (in)equivalence of the translation, and finally instructions to repair the translation if it is not equivalent.  We use Claude Code~\cite{claudecode} as the agent for most of our experiments, which comes with a set of tools out of the box, namely, reading + writing files, executing arbitrary shell commands, and searching the web. The agent outputs an overall equivalence verdict, a set of tests in both the source and target PL, and a translation patch if the agent believed the translation was not equivalent.

\subsection{Verdict Agent}
\label{subsec:verdict}

The final component of \approach is the Verdict Agent, which produces a definitive assessment of the translation’s correctness by synthesizing the information from the previous two stages. The Verdict Agent takes as input the semantic analysis report and the test execution + repair report. It leverages another LLM agent to consolidate these results into a final verdict. This agent's primary job is to (1) confirm the results of the Test Generator \& Repair Agent, and (2) to produce a condensed summary of the results, which is useful for end-users.
\section{Evaluation}
\label{sec:evaluation}

We conducted an extensive set of empirical studies, evaluating the effectiveness of \approach in translation validation (\S\ref{subsec:rq1}), translation repair (\S\ref{subsec:rq2}), investigation of development cost and adaptability (\S\ref{subsec:rq3}), and ablation studies to justify the architecture of \approach (\S\ref{subsec:rq4}). We evaluate \approach on benchmarks used in prior work~\cite{ibrahimzada2025alphatrans,wang2025skel,zhang2025oxidizer,ou2024repository} and compare against them in translation validation and repair. For LLM and agentic framework, we used \texttt{\small Claude 3.7 Sonnet}~\cite{claude} and Claude Code~\cite{claudecode}. More experimental setup details are listed in Appendix~\ref{appendix:experiment-setup}.








\subsection{Effectiveness in Translation Validation}
\label{subsec:rq1}

\begin{table*}[t]
    \scriptsize
    \centering
    \caption{Effectiveness of \approach in translation validation compared to existing techniques. \textbf{EQ:}~Equivalent, \textbf{NEQ:} Not Equivalent, \textbf{VF:} Validation Failure, \textbf{Agreement}: number and percentage of translation pairs where \approach's and the existing tool's verdicts agree, \textbf{Disagreement}: percentage of disagreements ruled in favor of \textbf{Tool} and \approach (\textbf{Ours}). \textbf{VF}s are excluded from \textbf{Agreement} and \textbf{Disagreement} calculations.}
    \vspace{-5pt}
    \begin{tabular}{c|lc|ccc|ccc|c|cc} 
\hline
\multicolumn{1}{l|}{\multirow{2}{*}{\textbf{Tool}}}                                      & \multirow{2}{*}{\textbf{Project}} & \multirow{2}{*}{\begin{tabular}[c]{@{}c@{}}\textbf{Total \#}\\\textbf{Trans. Pairs}\end{tabular}} & \multicolumn{3}{c|}{\textbf{Tool Validation }}                                             & \multicolumn{3}{c|}{\textbf{\approach}}                                                  & \multirow{2}{*}{\textbf{Agreement}} & \multicolumn{2}{c}{\textbf{Disagreement}}           \\ 
\cline{4-9}\cline{11-12}
\multicolumn{1}{l|}{}                                                                    &                                   &                                                                                                   & \multicolumn{1}{c|}{\textbf{EQ}} & \multicolumn{1}{c|}{\textbf{NEQ}} & \textbf{VF}         & \multicolumn{1}{c|}{\textbf{EQ}} & \multicolumn{1}{c|}{\textbf{NEQ}} & \textbf{VF}       &                                     & \multicolumn{1}{c|}{\textbf{Tool}} & \textbf{Ours}  \\ 
\hline
\multirow{6}{*}{\rotatebox{90}{\oxidizer}}                                                & checkdigit                        & 29                                                                                                & 21 (72.4)                        & 8 (27.6)                          & 0 (0)               & 22 (75.9)                        & 7 (24.1)                          & 0 (0)             & 24 (82.8)                           & 0.0                                & 100            \\
                                                                                         & go-edlib                          & 24                                                                                                & 18 (75)                          & 6 (25)                            & 0 (0)               & 16 (66.7)                        & 8 (33.3)                          & 0 (0)             & 14 (58.3)                           & 11.1                               & 88.9           \\
                                                                                         & histogram                         & 19                                                                                                & 12 (63.2)                        & 7 (36.8)                          & 0 (0)               & 11 (57.9)                        & 7 (36.8)                          & 1 (5.3)           & 8 (44.4)                            & 20.0                               & 80.0           \\
                                                                                         & nameparts                         & 15                                                                                                & 9 (60)                           & 6 (40)                            & 0 (0)               & 12 (80)                          & 3 (20)                            & 0 (0)             & 12 (80)                             & 33.3                               & 66.7           \\
                                                                                         & stats                             & 53                                                                                                & 38 (71.7)                        & 14 (26.4)                         & 1 (1.9)             & 37 (69.8)                        & 16 (30.2)                         & 0 (0)             & 35 (67.3)                           & 0.0                                & 100            \\
                                                                                         & textrank                          & 52                                                                                                & 40 (76.9)                        & 12 (23.1)                         & 0 (0)               & 34 (65.4)                        & 18 (34.6)                         & 0 (0)             & 28 (53.8)                           & 42.9                               & 57.1           \\ 
\hline
\rowcolor[rgb]{0.8,0.902,0.902} \multicolumn{1}{l|}{\textbf{Total}}                      &                                   & \textbf{192}                                                                                      & \textbf{138 (71.9)}              & \textbf{53 (27.6)}                & \textbf{1 (0.5)}    & \textbf{132 (68.8)}              & \textbf{59 (30.7)}                & \textbf{1 (0.5)}  & \textbf{121 (63.7)}                 & \textbf{15.9}                      & \textbf{84.1}  \\ 
\hline
\multirow{4}{*}{\rotatebox{90}{\begin{tabular}[c]{@{}c@{}}\textsc{Alpha}\\ \textsc{Trans}\end{tabular}}}    & cli                               & 273                                                                                               & 210 (76.9)                       & 24 (8.8)                          & 39 (14.3)           & 210 (76.9)                       & 60 (22)                           & 3 (1.1)           & 176 (76.2)                          & 30.0                               & 70.0           \\
                                                                                         & csv                               & 235                                                                                               & 97 (41.3)                        & 61 (26)                           & 77 (32.8)           & 185 (78.7)                       & 49 (20.9)                         & 1 (0.4)           & 108 (68.8)                          & 30.0                               & 70.0           \\
                                                                                         & fileupload                        & 192                                                                                               & 19 (9.9)                         & 1 (0.5)                           & 172 (89.6)          & 144 (75)                         & 48 (25)                           & 0 (0)             & 16 (80)                             & 25.0                               & 75.0           \\
                                                                                         & validator                         & 646                                                                                               & 247 (38.2)                       & 103 (15.9)                        & 296 (45.8)          & 483 (74.8)                       & 163 (25.2)                        & 0 (0)             & 225 (64.3)                          & 20.0                               & 80.0           \\ 
\hline
\rowcolor[rgb]{0.8,0.902,0.902} \multicolumn{1}{l|}{\textbf{Total}}                      &                                   & \textbf{1346}                                                                                     & \textbf{573 (42.6)}              & \textbf{189 (14)}                 & \textbf{584 (43.4)} & \textbf{1022 (75.9)}             & \textbf{320 (23.8)}               & \textbf{4 (0.3)}  & \textbf{525 (69.3)}                 & \textbf{26.5}                      & \textbf{73.5}  \\ 
\hline
\multirow{8}{*}{\rotatebox{90}{\skel}}                                                    & bst                               & 19                                                                                                & 19 (100)                         & 0 (0)                             & 0 (0)               & 14 (73.7)                        & 5 (26.3)                          & 0 (0)             & 14 (73.7)                           & 20.0                               & 80.0           \\
                                                                                         & colorsys                          & 8                                                                                                 & 8 (100)                          & 0 (0)                             & 0 (0)               & 7 (87.5)                         & 1 (12.5)                          & 0 (0)             & 7 (87.5)                            & 0.0                                & 100            \\
                                                                                         & heapq                             & 22                                                                                                & 19 (86.4)                        & 3 (13.6)                          & 0 (0)               & 12 (54.5)                        & 10 (45.5)                         & 0 (0)             & 13 (59.1)                           & 50.0                               & 50.0           \\
                                                                                         & html                              & 44                                                                                                & 40 (90.9)                        & 2 (4.5)                           & 2 (4.5)             & 35 (79.5)                        & 9 (20.5)                          & 0 (0)             & 33 (78.6)                           & 66.7                               & 33.3           \\
                                                                                         & mathgen                           & 81                                                                                                & 77 (95.1)                        & 4 (4.9)                           & 0 (0)               & 65 (80.2)                        & 16 (19.8)                         & 0 (0)             & 67 (82.7)                           & 50.0                               & 50.0           \\
                                                                                         & rbt                               & 27                                                                                                & 26 (96.3)                        & 0 (0)                             & 1 (3.7)             & 23 (85.2)                        & 4 (14.8)                          & 0 (0)             & 22 (84.6)                           & 75.0                               & 25.0           \\
                                                                                         & strsim                            & 64                                                                                                & 50 (78.1)                        & 0 (0)                             & 14 (21.9)           & 56 (87.5)                        & 8 (12.5)                          & 0 (0)             & 44 (88)                             & 40.0                               & 60.0           \\
                                                                                         & toml                              & 72                                                                                                & 37 (51.4)                        & 10 (13.9)                         & 25 (34.7)           & 49 (68.1)                        & 22 (30.6)                         & 1 (1.4)           & 33 (71.7)                           & 40.0                               & 60.0           \\ 
\hline
\rowcolor[rgb]{0.8,0.902,0.902} \multicolumn{1}{l|}{\textbf{Total}}                      &                                   & \textbf{337}                                                                                      & \textbf{276 (81.9)}              & \textbf{19 (5.6)}                 & \textbf{42 (12.5)}  & \textbf{261 (77.4)}              & \textbf{75 (22.3)}                & \textbf{1 (0.3)}  & \textbf{233 (79.3)}                 & \textbf{46.5}                      & \textbf{53.5}  \\ 
\hline
\multirow{6}{*}{\rotatebox{90}{\begin{tabular}[c]{@{}c@{}}\textsc{RustRepo}\\ \textsc{Trans}\end{tabular}}} & charset                           & 33                                                                                                & 20 (60.6)                        & 13 (39.4)                         & 0 (0)               & 14 (42.4)                        & 19 (57.6)                         & 0 (0)             & 25 (75.8)                           & 75.0                               & 25.0           \\
                                                                                         & deltachat                         & 125                                                                                               & 54 (43.2)                        & 69 (55.2)                         & 2 (1.6)             & 39 (31.2)                        & 84 (67.2)                         & 2 (1.6)           & 92 (76)                             & 90.0                               & 10.0           \\
                                                                                         & iceberg-java                      & 25                                                                                                & 9 (36)                           & 16 (64)                           & 0 (0)               & 3 (12)                           & 16 (64)                           & 6 (24)            & 15 (78.9)                           & 100                                & 0.0            \\
                                                                                         & iceberg-py                        & 44                                                                                                & 15 (34.1)                        & 28 (63.6)                         & 1 (2.3)             & 11 (25)                          & 29 (65.9)                         & 4 (9.1)           & 35 (89.7)                           & 100                                & 0.0            \\
                                                                                         & crypto-c                          & 20                                                                                                & 16 (80)                          & 4 (20)                            & 0 (0)               & 7 (35)                           & 13 (65)                           & 0 (0)             & 11 (55)                             & 66.7                               & 33.3           \\
                                                                                         & crypto-java                       & 97                                                                                                & 39 (40.2)                        & 58 (59.8)                         & 0 (0)               & 30 (30.9)                        & 67 (69.1)                         & 0 (0)             & 86 (88.7)                           & 100                                & 0.0            \\ 
\hline
\rowcolor[rgb]{0.8,0.902,0.902} \multicolumn{1}{l|}{\textbf{Total}}                      &                                   & \textbf{344}                                                                                      & \textbf{153 (44.5)}              & \textbf{188 (54.7)}               & \textbf{3 (0.9)}    & \textbf{104 (30.2)}              & \textbf{228 (66.3)}               & \textbf{12 (3.5)} & \textbf{264 (80.2)}                 & \textbf{87.5}                      & \textbf{12.5}  \\ 
\hline\hline
\rowcolor[rgb]{0.902,0.902,0.902} \multicolumn{1}{l|}{\textbf{Total}}                    &                                   & \textbf{2219}                                                                                     & \textbf{1140 (51.4)}             & \textbf{449 (20.2)}               & \textbf{630 (28.4)} & \textbf{1519 (68.5)}             & \textbf{682 (30.7)}               & \textbf{18 (0.8)} & \textbf{1143 (72.8)}                & \textbf{39.3}                      & \textbf{60.7}  \\
\hline
\end{tabular}
    \vspace{-10pt}
    \label{table:rq1-validation}
\end{table*}

To assess effectiveness in translation validation, we run \approach on each translation pair to obtain an equivalence verdict, and compare it with the verdict of existing tools. The columns under \textbf{Tool Validation} and \textbf{\approach} in Table~\ref{table:rq1-validation} summarize the equivalence verdicts for the competing validation tool and \approach, respectively. There are three types of equivalence verdicts: (1) Equivalent (\textbf{EQ}) indicating the source function and its translation are equivalent, (2) Not Equivalent (\textbf{NEQ}) indicating they are inequivalent, and (3) Validation Failure (\textbf{VF}) indicating that the tool failed to provide a verdict. 
Competing tools can fail to provide a verdict if
(1) the source project did not have unit tests covering the function, 
or (2) the competing tool's language interoperability mechanism crashes before providing verdict. 
\approach can fail to provide a verdict if the timeout limit is reached. The \textbf{Agreement} column shows the number and percentage where both \approach's and others verdicts agree or disagree. The \textbf{Disagreement} column shows the result of our human investigation on disagreements. The \textbf{Tool} sub-column shows the percentage of disagreements where the existing tool's verdict was correct, and \textbf{Ours} sub-column shows the same metric for \approach. On average, \approach takes $309$ seconds to produce a verdict, with an average cost of \$$1.22$ and a total cost of \$$2{,}710.45$. 

These results demonstrate effectiveness of \approach in providing an equivalence verdict: \approach provides verdicts for $99.7\%$ of translation pairs from \alphatrans and \skel, whereas these techniques report verdicts for only $56.6\%$ and $87.5\%$ of their studied translation pairs, respectively. In addition, \approach's verdicts show a high level of agreement with all prior work, ranging from $63.7\%$ to $80.2\%$. Furthermore, we manually investigate disagreements (\S\ref{subsec:rq1-dispute-analysis}) and show examples of incorrect verdicts (\S\ref{subsec:rq1-incorrect-verdicts}).

\subsection{Effectiveness in Translation Repair}
\label{subsec:rq2}

\begin{table*}[t]
    \scriptsize
    \centering
    \caption{Effectiveness of \approach in translation repair compared against existing techniques. \textbf{NEQ:} Not Equivalent, \textbf{NR:} Not Reported/Repaired.}
    \vspace{-5pt}
    \begin{tabular}{c|lc|c|c|c|c|c} 
\hline
\textbf{Tool}                                                                                               & \textbf{Project}     & \begin{tabular}[c]{@{}c@{}}\textbf{Total \#}\\\textbf{Trans. Pair}\end{tabular} & \begin{tabular}[c]{@{}c@{}}\textbf{Tool NEQ~$\cap$}\\\textbf{\approach NEQ}\end{tabular} & \begin{tabular}[c]{@{}c@{}}\textbf{Tool}\\\textbf{Repaired}\end{tabular} & \begin{tabular}[c]{@{}c@{}}\textbf{\approach}\\\textbf{\textbf{Repaired}}\end{tabular} & \begin{tabular}[c]{@{}c@{}}\textbf{Disagreement}\\\textbf{Repaired}\end{tabular} & \begin{tabular}[c]{@{}c@{}}\textbf{Coverage}\\\textbf{(Improvement)~\%}\end{tabular}  \\ 
\hline
\multirow{6}{*}{\rotatebox{90}{\oxidizer}}                                                                  & checkdigit           & 29                                                                              & 5 (17.2)                                                                                 & NR                                                                       & 5 (100)                                                                                & 2 (100)                                                                          & 86.2 (0)                                                                              \\
                                                                                                            & go-edlib             & 24                                                                              & 2 (8.3)                                                                                  & NR                                                                       & 1 (50)                                                                                 & 4 (100)                                                                          & 100 (0)                                                                               \\
                                                                                                            & histogram            & 19                                                                              & 2 (10.5)                                                                                 & NR                                                                       & 1 (50)                                                                                 & 2 (66.7)                                                                         & 68.4 (0)                                                                              \\
                                                                                                            & nameparts            & 15                                                                              & 3 (20)                                                                                   & NR                                                                       & 1 (33.3)                                                                               & 0 (0)                                                                            & 100 (0)                                                                               \\
                                                                                                            & stats                & 53                                                                              & 6 (11.3)                                                                                 & NR                                                                       & 6 (100)                                                                                & 5 (100)                                                                          & 100 (\textbf{\textbf{\textbf{\textbf{$\uparrow$}}}1.9})                               \\
                                                                                                            & textrank             & 52                                                                              & 3 (5.8)                                                                                  & NR                                                                       & 3 (100)                                                                                & 2 (100)                                                                          & 100 (0)                                                                               \\ 
\hline
\rowcolor[rgb]{0.8,0.902,0.902} \textbf{Total}                                                              & \multicolumn{1}{c}{} & \textbf{192}                                                                    & \textbf{21 (10.9)}                                                                       & \textbf{\textbf{0 (0)}}                                                  & \textbf{17 (81)}                                                                       & \textbf{15 (93.8)}                                                               & \textbf{92.4 ($\uparrow$0.3)}                                                         \\ 
\hline
\multirow{4}{*}{\rotatebox{90}{\begin{tabular}[c]{@{}c@{}}\textsc{Alpha}\\ \textsc{Trans}\end{tabular}}}    & cli                  & 273                                                                             & 9 (3.3)                                                                                  & NR                                                                       & 7 (77.8)                                                                               & 3 (100)                                                                          & 100 (\textbf{\textbf{$\uparrow$}6.2})                                                 \\
                                                                                                            & csv                  & 235                                                                             & 20 (8.5)                                                                                 & NR                                                                       & 16 (80)                                                                                & 2 (100)                                                                          & 100 (\textbf{\textbf{$\uparrow$}11.9})                                                \\
                                                                                                            & fileupload           & 192                                                                             & 0 (0)                                                                                    & -                                                                        & -                                                                                      & 2 (100)                                                                          & 98.9 (\textbf{\textbf{$\uparrow$}78.1})                                               \\
                                                                                                            & validator            & 646                                                                             & 34 (5.3)                                                                                 & NR                                                                       & 23 (67.6)                                                                              & 3 (100)                                                                          & 99.3 (\textbf{\textbf{$\uparrow$}36.8})                                               \\ 
\hline
\rowcolor[rgb]{0.8,0.902,0.902} \textbf{Total}                                                              & \multicolumn{1}{c}{} & \textbf{1346}                                                                   & \textbf{63 (4.7)}                                                                        & \textbf{0 (0)}                                                           & \textbf{46 (73)}                                                                       & \textbf{10 (100)}                                                                & \textbf{99.6 ($\uparrow$33.3)}                                                        \\ 
\hline
\multirow{8}{*}{\rotatebox{90}{\skel}}                                                                      & bst                  & 19                                                                              & 0 (0)                                                                                    & -                                                                        & -                                                                                      & 4 (100)                                                                          & 100 (0)                                                                               \\
                                                                                                            & colorsys             & 8                                                                               & 0 (0)                                                                                    & -                                                                        & -                                                                                      & 1 (100)                                                                          & 100 (0)                                                                               \\
                                                                                                            & heapq                & 22                                                                              & 2 (9.1)                                                                                  & NR                                                                       & 1 (50)                                                                                 & 3 (100)                                                                          & 100 (0)                                                                               \\
                                                                                                            & html                 & 44                                                                              & 1 (2.3)                                                                                  & NR                                                                       & 0 (0)                                                                                  & 2 (100)                                                                          & 100 (\textbf{\textbf{\textbf{\textbf{$\uparrow$}}}4.5})                               \\
                                                                                                            & mathgen              & 81                                                                              & 3 (3.7)                                                                                  & NR                                                                       & 1 (33.3)                                                                               & 2 (66.7)                                                                         & 100 (0)                                                                               \\
                                                                                                            & rbt                  & 27                                                                              & 0 (0)                                                                                    & -                                                                        & -                                                                                      & 1 (100)                                                                          & 100 (\textbf{\textbf{\textbf{\textbf{$\uparrow$}}}3.7})                               \\
                                                                                                            & strsim               & 64                                                                              & 0 (0)                                                                                    & -                                                                        & -                                                                                      & 3 (100)                                                                          & 100 (\textbf{\textbf{\textbf{\textbf{$\uparrow$}}}21.9})                              \\
                                                                                                            & toml                 & 72                                                                              & 5 (6.9)                                                                                  & NR                                                                       & 3 (60)                                                                                 & 3 (100)                                                                          & 100 (\textbf{\textbf{\textbf{\textbf{$\uparrow$}}}34.7})                              \\ 
\hline
\rowcolor[rgb]{0.8,0.902,0.902} \textbf{Total}                                                              & \multicolumn{1}{c}{} & \textbf{337}                                                                    & \textbf{11 (3.3)}                                                                        & \textbf{0 (0)}                                                           & \textbf{5 (45.5)}                                                                      & \textbf{19 (95)}                                                                 & \textbf{100 ($\uparrow$8.1)}                                                          \\ 
\hline
\multirow{6}{*}{\rotatebox{90}{\begin{tabular}[c]{@{}c@{}}\textsc{RustRepo}\\ \textsc{Trans}\end{tabular}}} & charset              & 33                                                                              & 12 (36.4)                                                                                & 5 (41.7)                                                                 & 7 (58.3)                                                                               & 1 (100)                                                                          & 100 (0)                                                                               \\
                                                                                                            & deltachat            & 125                                                                             & 60 (48)                                                                                  & 10 (16.7)                                                                & 11 (18.3)                                                                              & 1 (100)                                                                          & 100 (\textbf{\textbf{$\uparrow$}1.6})                                                 \\
                                                                                                            & iceberg-java         & 25                                                                              & 12 (48)                                                                                  & 1 (8.3)                                                                  & 1 (8.3)                                                                                & 0 (0)                                                                            & 100 (0)                                                                               \\
                                                                                                            & iceberg-py           & 44                                                                              & 25 (56.8)                                                                                & 4 (16)                                                                   & 6 (24)                                                                                 & 0 (0)                                                                            & 100 (\textbf{\textbf{$\uparrow$}2.3})                                                 \\
                                                                                                            & crypto-c             & 20                                                                              & 4 (20)                                                                                   & 1 (25)                                                                   & 2 (50)                                                                                 & 1 (100)                                                                          & 100 (0)                                                                               \\
                                                                                                            & crypto-java          & 97                                                                              & 57 (58.8)                                                                                & 28 (49.1)                                                                & 39 (68.4)                                                                              & 0 (0)                                                                            & 100 (0)                                                                               \\ 
\hline
\rowcolor[rgb]{0.8,0.902,0.902} \textbf{Total}                                                              & \multicolumn{1}{c}{} & \textbf{344}                                                                    & \textbf{170 (49.4)}                                                                      & \textbf{49 (28.8)}                                                       & \textbf{66 (38.8)}                                                                     & \textbf{3 (100)}                                                                 & \textbf{100 ($\uparrow$0.6)}                                                          \\ 
\hline\hline
\rowcolor[rgb]{0.902,0.902,0.902} \textbf{\textbf{Total}}                                                   & \multicolumn{1}{c}{} & \textbf{2219}                                                                   & \textbf{265 (11.9)}                                                                      & \textbf{49 (18.5)}                                                       & \textbf{134 (50.6)}                                                                    & \textbf{47 (95.9)}                                                               & \textbf{98.1 (\textbf{$\uparrow$~}8.5)}                                               \\
\hline
\end{tabular}
    \vspace{-5pt}
    \label{table:rq2-repair}
\end{table*}

We investigate the effectiveness of \approach in automated repair of translation bugs, and its ability to improve code coverage of existing projects. Table~\ref{table:rq2-repair} shows the results of this research question. To fairly evaluate the effectiveness of existing tools and \approach in translation repair, we extract a subset of translations where both techniques generated a patch. 

In total, $\frac{265}{2219}$ ($11.9\%$) buggy translations were considered for our study. To validate patches, we used original project tests that previously failed on buggy translations, and only considered a patch correct when all failing tests passed. We did not use generated tests by \approach to avoid bias in our evaluation. Column \textit{Tool Repaired} shows the number of buggy translations repaired by existing techniques. Only $\frac{49}{265}$ ($18.5\%$) bugs have been repaired by prior techniques. Except for \rustrepotrans~\cite{ou2024repository}, other code translation tools failed to generate correct patches to repair translation bugs. \skel~\cite{wang2025skel} reprompts the same LLM for repairing bugs, but then requires a user to manually provide a fix. \alphatrans~\cite{ibrahimzada2025alphatrans} and \oxidizer~\cite{zhang2025oxidizer} generates patches in the loop, however, no effectiveness was reported in their papers, and we could not repair any translation bugs using their tools. Moreover, patches by \alphatrans and \oxidizer could not be validated mostly because of limitations in their validation system. For example, the following code snippets show instance \texttt{\small 11} from the project \texttt{\small commons-validator} in \alphatrans. The GraalVM-based validation system in \alphatrans does not validate this translation as functionally equivalent, although our manual investigation indicates that the \texttt{\small Python} translation is correct. Therefore, the limitation in \alphatrans is mostly due to its validation system being unable to validate LLM patches.

\vspace{-10pt}
\begin{minted}[frame=lines,framesep=1mm,baselinestretch=0.5, fontsize=\scriptsize, breaklines, breakanywhere, linenos,numbersep=2pt]{java}
------------------- JAVA SOURCE CODE ------------------
public boolean isOn(long flag) {
    return (this.flags & flag) == flag;
}
\end{minted}
\vspace{-23pt}
\begin{minted}[escapeinside=||, frame=lines,framesep=1mm,baselinestretch=0.5, fontsize=\scriptsize, breaklines, breakanywhere, linenos,numbersep=2pt]{py}
------------------ PYTHON TRANSLATION -----------------
def isOn(self, flag: int) -> bool:

    return (self.__flags & flag) == flag
\end{minted}
\vspace{-10pt}

Column \textit{\approach Repaired} shows the number of translation bugs successfully repaired by our approach. In total, \approach repaired $\frac{134}{265}$ ($50.6\%$) of translation bugs, $32.1\%$ more than existing reprompting-based techniques. Given the limitations in validation system of \alphatrans discussed earlier, we manually investigate and validate patches from this tool
\footnote{Please refer to our artifacts website~\cite{website} for detailed investigation results.}.
The following code snippets demonstrate instance \texttt{\small 276} from the project \texttt{\small commons-validator} in \alphatrans in which its reprompting-based repairing could not generate a correct patch. By contrast, \approach successfully repairs this translation bug with the help of its Library Analyzer (\S\ref{subsubsec:library}). The report generated by this semantic analyzer indicate \texttt{\small "... the standard Python datetime.date class does not have a SHORT attribute ..."} which is correct. The Test Generator \& Repair Agent (\S\ref{subsec:test-gen-repair}) then leverages this analysis and successfully generate a patch by replacing \texttt{\small SHORT} with constant \texttt{\small 3}.

\vspace{-10pt}
\begin{minted}[frame=lines,framesep=1mm,baselinestretch=0.5, fontsize=\scriptsize, breaklines, breakanywhere, linenos,numbersep=2pt]{java}
------------------- JAVA SOURCE CODE ------------------
public static DateValidator DateValidator1() {
    return new DateValidator(true, DateFormat.SHORT);
}
\end{minted}
\vspace{-23pt}
\begin{minted}[escapeinside=||, frame=lines,framesep=1mm,baselinestretch=0.5, fontsize=\scriptsize, breaklines, breakanywhere, linenos,numbersep=2pt, highlightlines={3,4}]{py}
------------------ PYTHON TRANSLATION -----------------
def DateValidator1() -> DateValidator:
|\xglobal\colorlet{FancyVerbHighlightColor}{carnationpink}|- return DateValidator(True, datetime.date.SHORT)
|\xglobal\colorlet{FancyVerbHighlightColor}{powderblue}|+ return DateValidator(True, 3)
\end{minted}
\vspace{-10pt}

Column \textit{Disagreement Repaired} shows the number of disagreements from RQ1 (\S\ref{subsec:rq1}) which \approach determined as \emph{not equivalent} and successfully generated a correct patch. Of the $49$ disagreements resolved in favor of \approach, we further asked our manual investigators if the generated patch was correct or not. On average, $\frac{47}{49}$ ($95.9\%$) of patches were validated as correct. Notice that this column cannot be directly compared with existing techniques, because they validated disagreements as functionally correct and did not generate any patches. Moreover, Column \textit{Coverage} indicates the overall coverage for subject projects and the total improvement as a result of tests generated by \approach. The generated tests help improve code coverage by $8.5\%$ (from $89.6\%$ to $98.1\%$). Project \texttt{\small commons-fileupload} from \alphatrans sees the most improvement ($78.1\%$), with most of its translated fragments now validated with \approach tests.


\subsection{Development Cost and Adaptability}
\label{subsec:rq3}



\begin{table}[t]
    \scriptsize
    \centering
    \caption{Development cost of \approach compared against existing tools.}
    \resizebox{\linewidth}{!}{
        \begin{tabular}{c|c|c} 
\hline
\textbf{Tool}    & \begin{tabular}[c]{@{}c@{}}\textbf{Complexity}\\\textbf{(LoC)}\end{tabular} & \begin{tabular}[c]{@{}c@{}}\textbf{Development Time}\\\textbf{(Months)}\end{tabular}  \\ 
\hline
\alphatrans      & $10859$                                                                     & 8                                                                                     \\
\oxidizer        & $19052$                                                                     & 10                                                                                    \\
\skel            & $3843$                                                                      & 6                                                                                     \\ 
\hdashline
\approach (ours) & $1650$                                                                      & 0.7                                                                                   \\
\hline
\end{tabular}
    }
    \label{table:dev-cost}
\end{table}

\subsubsection{Development Cost}
\label{subsubsec:rq3-cost}

Table~\ref{table:dev-cost} shows the development cost of \approach against existing techniques. We define cost as the tool's total lines of code (LoC). As shown in the table, developing \approach is cheap and the initial version only consists of $1{,}650$ LoC, supporting $6$ different PLs. Its dependence only on the Tree-Sitter~\cite{tree-sitter} parser makes it easy to support more languages using only $280$ LoC. By contrast, other tools, such as, \skel~\cite{wang2025skel}, \alphatrans~\cite{ibrahimzada2025alphatrans}, and \oxidizer~\cite{zhang2025oxidizer} only support \emph{one} PL pair and require significant engineering effort to adapt to more languages. More precisely, \approach is $\times2.3$, $\times6.6$, and $\times11.6$ cheaper than \skel, \alphatrans, and \oxidizer, respectively. The static nature of \approach makes it cost-effective and scalable, achieving better performance and revealing major limitations in existing tools.

\subsubsection{Adaptability}
\label{subsubsec:rq3-adaptability}

The architecture of \approach is largely independent of any specific LLM or agent framework, making it easy to extend and integrate with more LLMs. In this research question, we investigate the extent to which replacing  Anthropic’s Claude 3.7 Sonnet with  OpenAI \texttt{\small o4-mini-2025-04-16}~\cite{o4mini}, and Claude Code with Codex~\cite{codex} agent, impacts the performance of \approach. 
Due to the limited cost budget, 
we randomly sampled $96$ instances from all subject projects. While sampling, we controlled for equal contribution of equivalent and non-equivalent translations, resulting in 
$58$ and $38$ samples for each, respectively. 

\begin{figure}
    \centering
    \includegraphics[width=0.7\linewidth]{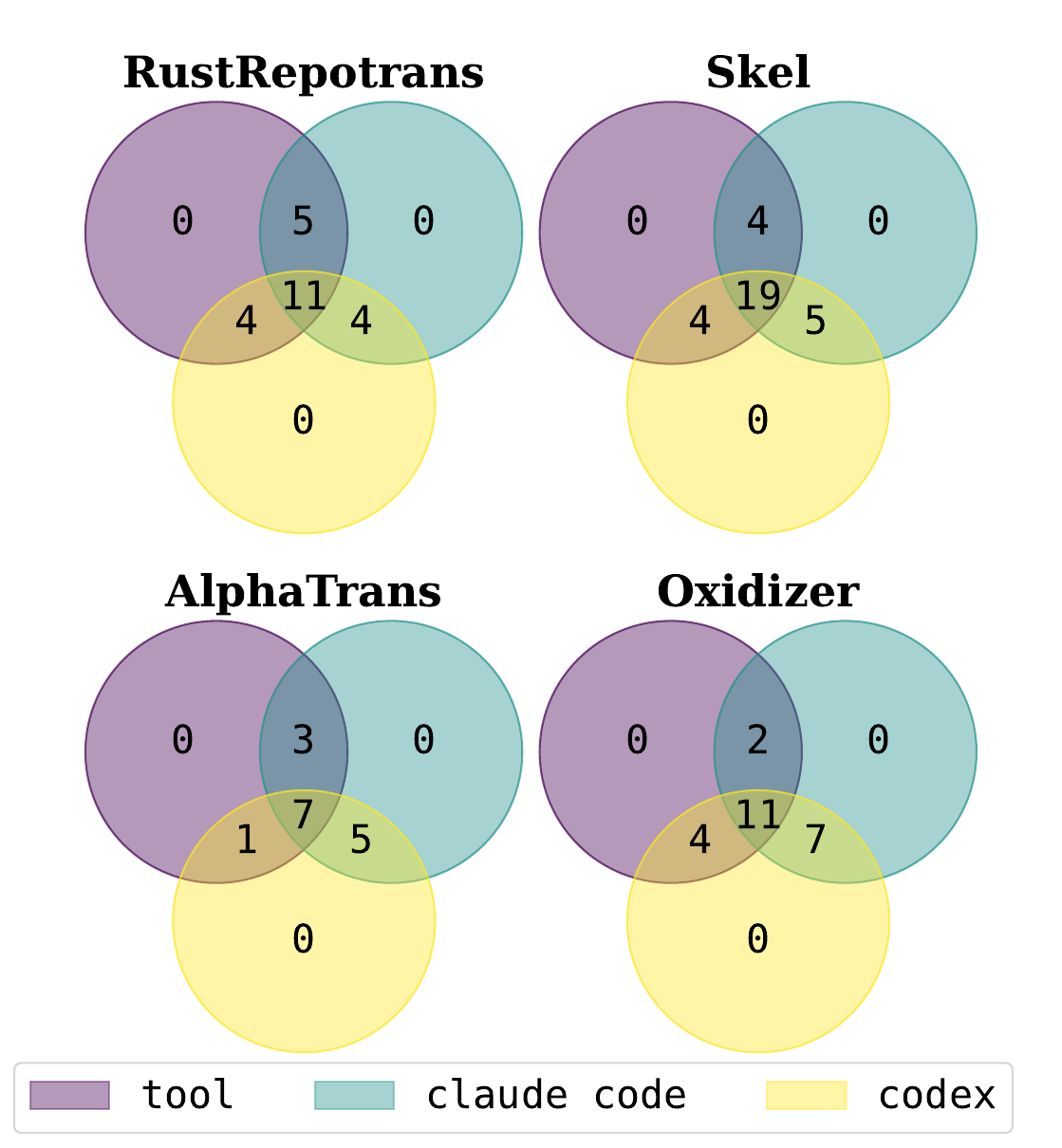}
    \caption{Agreement and dispute cases between tool validation system and \approach with Claude and Codex agents.}
    \label{fig:venn-diagram}
    \vspace{-10pt}
\end{figure}

Figure~\ref{fig:venn-diagram} illustrates the results of this study. Our analysis indicates that \approach with Claude Code and Codex agrees $73\%$ of the time against existing validation systems. Moreover, we also analyzed both agents' behavior in terms of problem understanding and finding a solution. Our investigation of agent trajectories (footprint of agent actions) shows OpenAI's Codex makes fewer actions to explore the codebase, and attempts to provide a decision faster. By contrast, Anthropic's Claude Code agent first plans and reasons thoroughly about its tasks, specifically called \texttt{TODO List} by the agent, and then takes specific actions to perform each of its planned tasks.



\subsection{Ablation Study}
\label{subsec:rq4}


\subsubsection{Impact of Semantic Analyzer and Test Generator Agent}
\label{subsubsec:rq4-impact-semantic-analyzer-test-gen-agent}

\begin{figure*}[ht]
    \centering
    \includegraphics[width=0.9\textwidth]{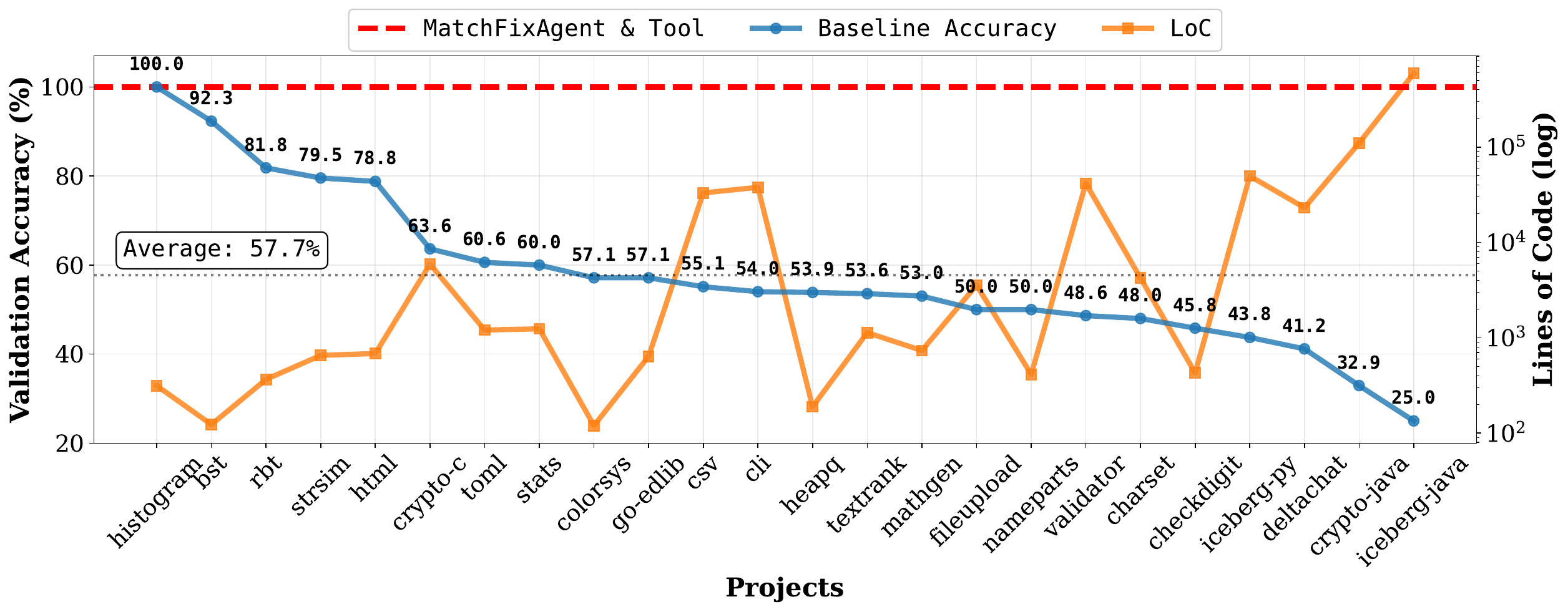}
    \caption{Impact of semantic analyzer and test generator agent in \approach. A standalone baseline agent struggles validating translations as the projects grow in size.}
    \label{fig:ablation-impact}
    \vspace{-15pt}
\end{figure*}

To investigate the importance of semantic analyzer and test generator agent in \approach, we evaluated a standalone baseline agent that uses the same LLM and agent framework, e.g., \texttt{\small Claude Sonnet 3.7}~\cite{claude} and Claude Code~\cite{claudecode}
\footnote{Please refer to our artifacts~\cite{website} for prompt templates used in this study.}.
In order to perform a controlled study, we created a sample from the original dataset where existing tools and \approach verdicts agree with each other, meaning we collected all non-dispute instances. In total, $1{,}091$ instances, $862$ equivalent, and $229$ non-equivalent translations were selected. Figure~\ref{fig:ablation-impact} shows the result of this ablation study. 
Accuracy indicates the ratio of baseline verdicts that agree with the existing tool and \approach. On average, the validation accuracy drops by $42.3\%$, illustrating the importance of \approach's semantic analyzer and test generator components. Across all projects, the baseline agent only reproduced \approach and tool results in the \texttt{\small histogram} project from \oxidizer~\cite{zhang2025oxidizer}, achieving $100\%$ validation correctness. For the remaining projects, the agent's accuracy dropped as low as $25\%$ in \texttt{\small iceberg-java}, which is the project with the largest number of lines of code.

Moreover, we also evaluated our baseline agent on dispute instances, a total of $416$ cases between \approach and existing tools. The results indicate $48.3\%$ agreement with \approach, and $41.1\%$ with competing tools, however, these numbers do not convey any important meaning without ground-truths.

To address this issue, we considered the $145$ cases where manual investigation established ground truth (\S\ref{subsec:rq1-dispute-analysis}), and we can directly measure baseline agent accuracy. On the $145$ manually resolved cases, the baseline agent achieves only $47.7\%$ accuracy on the $88$ cases where \approach is correct--the cases that represent \approach's unique advantage over existing tools. By contrast, the baseline achieves $56.1\%$ on the $57$ cases where the existing tool is correct, confirming that the baseline agent systematically fails on precisely the hard, ambiguous translations that \approach’s semantic analyzer and test generation are designed to handle.

The $47.7\%$ accuracy on difficult, disputed cases (where our original ablation showed only a $42.3\%$ drop on easier, non-disputed cases) confirms that \approach's multi-agent architecture is essential, not just helpful. The baseline agent's near-random performance on these difficult cases demonstrates that the semantic analyzer and test generation components are the core drivers of \approach's accuracy advantage in resolving both straightforward and challenging translations.


\subsubsection{Impact of Semantic Analyzer}
\label{subsubsec:rq4-impact-semantic-analyzer}

\begin{figure}
    \centering
    \includegraphics[width=0.7\linewidth]{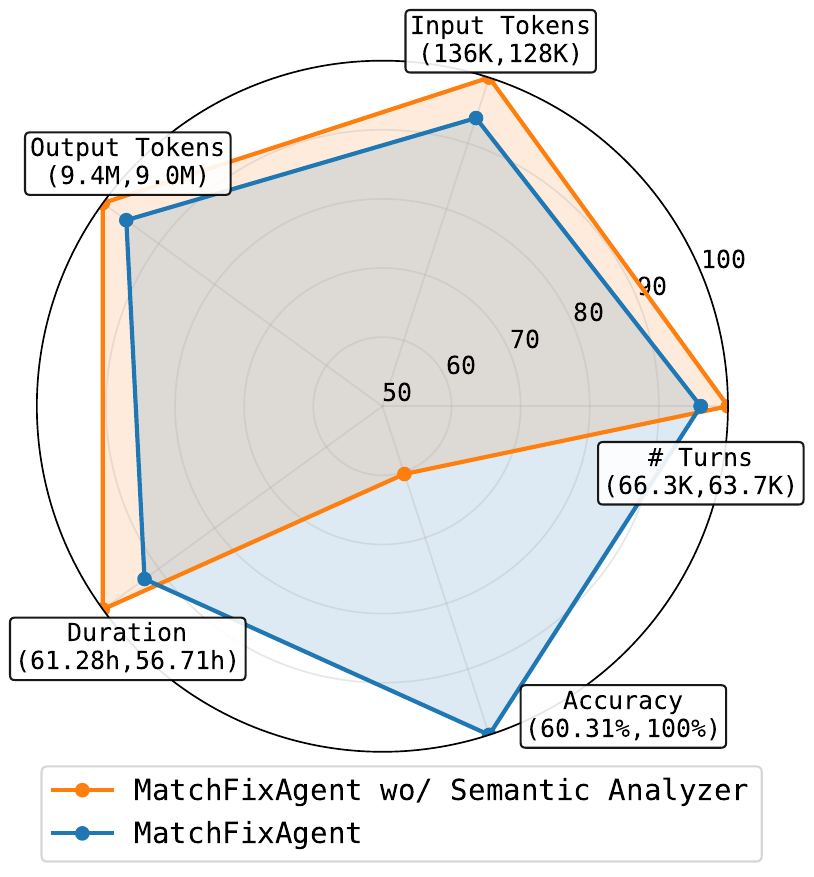}
    \caption{Removing the semantic analyzer decreases the effectiveness of \approach, while increasing token consumption, number of turns, and processing time.}
    \label{fig:radar-chart}
\end{figure}

We perform another study by removing only the semantic analysis results when prompting the test generator and verdict agents. We use the same set with $1{,}091$ instances from the previous ablation to perform this study. Figure~\ref{fig:radar-chart} illustrates the result of our second ablation. The results indicate that the performance of \approach significantly drops by $39.7\%$ without the six semantic analyses. Furthermore, we observe that the test generator agent without semantic analyzer is more costly and on average spends $3.9\%$ ($66.3K$ instead of $63.7K$), $6.2\%$ ($136K$ instead of $128K$), $4.2\%$ ($9.4M$ instead of $9.0M$), and $7.5\%$ ($61.28\,\mathrm{h}$ instead of $56.71\,\mathrm{h}$) more turns/interactions, input tokens, output tokens, and time, respectively. 


\section{Related Work}
\label{sec:related-work}

\textbf{Translation Validation and Repair.}
Existing automated translation validation techniques either rely on test
execution~\cite{pan2024lost, zhang2025oxidizer, ibrahimzada2025alphatrans,
shetty2024syzygy, ou2024repository, ziftci2025migrating, xue2025classeval,
yang2024exploring,ke2025advancing,dehghan2025translating,ibrahimzada2026recodeagent,roziere2020unsupervised,khatry2025crust}, formal methods~\cite{yang2025vert,nitin2024spectra,garzella2020leveraging},
or fuzzing~\cite{eniser2024towards}
for translation validation. Abid et al.~\cite{abid2024gluetest} and
\alphatrans~\cite{ibrahimzada2025alphatrans} leverage GraalVM~\cite{graalvm}
and language interoperability to execute code in both source and target PL
for translation validation. Other tools like
\oxidizer~\cite{zhang2025oxidizer} and \syzygy~\cite{shetty2024syzygy}
instrument programs to extract input-output pairs and use in target PL for
validation. \skel~\cite{wang2025skel} validates translations through test
execution by converting Python tests to JavaScript simply through string
replacement. This suffices since source Python tests are simple function
calls and value comparisons. For automated translation repair, most
tools~\cite{zhang2025oxidizer, wang2025skel, ibrahimzada2025alphatrans,
shetty2024syzygy, ou2024repository, yang2024exploring, pan2024lost} rely on
simple reprompting of LLMs with execution feedback, which has proven
ineffective.  Specifically, \alphatrans~\cite{ibrahimzada2025alphatrans}
performs independent reprompting of suspicious fragments based on execution
trace, while \skel~\cite{wang2025skel} requires a user for manually
repairing translation bugs.

\noindent \textbf{LLM Agents.} With the increasing prominence of agent-based
frameworks,
recent research and industrial
efforts have turned towards leveraging these frameworks to address various software engineering tasks.
SWE-agent~\cite{yang2024sweagent}
introduces a specialized agent-computer interface (ACI), facilitating agent
interaction with code repositories via file reading, editing, and execution
of bash commands.  \textsc{AutoCodeRover}~\cite{zhang2024autocoderover},
which provides LLM agents with specialized code-search APIs, enables
iterative retrieval and localization of code segments associated with bugs. 
\textsc{SpecRover}~\cite{ruan2025specrover} enhances \textsc{AutoCodeRover}
by emphasizing specification inference, generating function summaries, and
providing targeted feedback at crucial agent execution stages. 
Besides these state-of-the-art frameworks, numerous
additional agent-based approaches exist both in
open-source~\cite{wei2025swe, ouyang2025repograph, bouzenia2024repairagent, moatless, aider} and commercial products~\cite{wang2025openhands, amazonq,
devin}.

\section{Conclusion}
\label{sec:conclusion}

We presented \approach, a \emph{language-agnostic} technique that combines the power of program analysis and LLM agents for autonomous repository-level code translation validation and repair. It performs various semantic analyses to systematically generate targeted tests, enabling demonstration of functional equivalence or detection of semantic bugs. Through rigorous evaluation on multiple benchmarks and different PL-pairs, we show the effectiveness of \approach. 
Our manual investigation of generated reports reveals significant limitation of existing techniques. To our knowledge, \approach is the first approach that can effectively validate and repair translations in repository-level across multiple PLs.

\section*{Acknowledgements}

We thank the cohort of Summer 2025 interns and mentors at AWS for their valuable feedback. We also thank the anonymous reviewers for their comments, which helped make this work stronger, and Professor Elsa Gunter for partially funding one of the authors during this research.

\section*{Impact Statement}

This work is motivated to boost large language model agents in repository-level code translation validation and repair across multiple programming languages. Software developers spend a significant amount of time migrating code and testing them. We believe our tool will significantly improve the developers’ experience in their day-to-day life. We envision two ways developers can use \approach in their workflow. First, stakeholders can first migrate their codebase, and run \approach to clarify further requirements and detect missing corner-case functionality. Second, \approach can be directly integrated with existing and new automated code translation tools to provide a second level of confidence in translation validation.

\bibliography{bibliography}
\bibliographystyle{icml2026}

\newpage

\appendix
\onecolumn

\section{Software and Data}
\label{sec:software-data}

The implementation of \approach, the results of our human study, the agent logs and trajectories, and the docker images required for reproducing the results presented in this paper are publicly available~\cite{website}.

\section{Limitations of Prior Work}
\label{sec:limitations}

\begin{figure}[h]
    \centering
    \includegraphics[width=0.9\textwidth]{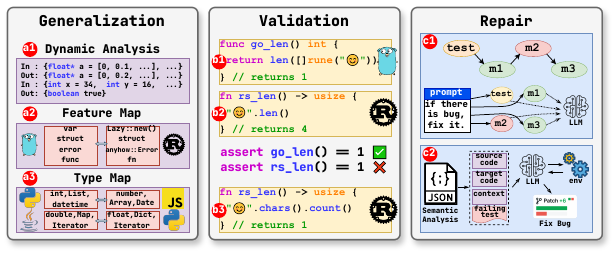}
    \vspace{-10pt}
    \caption{Illustration of key limitations of existing techniques in validation and repair of repository-level code translation and \approach addressing them.}
    \vspace{-10pt}
    \label{fig:limitations}
\end{figure}
To demonstrate the limitations of existing techniques~\cite{ibrahimzada2025alphatrans,zhang2025oxidizer,wang2025skel,shetty2024syzygy} for validation and repair, we use the examples in Figure~\ref{fig:limitations}. 

\textbf{Limitation 1: Generalization.}
Existing automated translation tools require substantial engineering effort to build validation systems for individual language pairs. For instance, \oxidizer~\cite{zhang2025oxidizer} and \syzygy~\cite{shetty2024syzygy} require dynamic analysis and I/O extraction from source code~(\circled[fill=red]{a1}). \oxidizer further requires a predefined feature map~(\circled[fill=red]{a2}), which together with their dynamic analyzer is $19{,}000$ lines of code. Other tools like \alphatrans~\cite{ibrahimzada2025alphatrans} and \skel~\cite{wang2025skel} validate programs using high quality type map~(\circled[fill=red]{a3}), which requires manual maintenance over time as PLs evolve. 
In comparison, \approach uses only language-agnostic LLM prompts, an off-the-shelf LLM coding agent tool, and lightweight static analysis. 
While the static analysis must be implemented for each PL, each PL requires approximately $280$ additional lines of code to support, hence it is extremely easy to generalize to many PLs.

\textbf{Limitation 2: Validation.}
Most prior work depends on existing source project tests for translation validation, but these tests may have insufficient coverage. 
The test suites in 
\alphatrans~\cite{ibrahimzada2025alphatrans} have an average of $56.57\%$ method coverage. Even when developer-written tests provide adequate coverage, they 
may miss 
the edge cases that expose subtle semantic differences between source and target languages. Consider the real-world case from \oxidizer~\cite{zhang2025oxidizer}, where a \texttt{\small Go} program~(\circled[fill=red]{b1}) that counts characters in a string is translated to \texttt{\small Rust}~(\circled[fill=red]{b2}) using the \texttt{\small .len()} method, which counts bytes rather than characters. \oxidizer validates this pair as \emph{functionally equivalent} because it only exercises this program with ASCII inputs. \approach, in contrast, marks the translation as not equivalent, and synthesizes a test with Unicode inputs (e.g., \texttt{\small U+1F60A}, a four-byte emoji representing a single character) to confirm the inequivalence. Upon detecting the translation bug, it 
automatically generates a patch that uses \texttt{\small .chars().count()} to ensure proper handling of both ASCII and Unicode characters~(\circled[fill=red]{b3}).

\textbf{Limitation 3: Repair.}
Current translation repair techniques solely rely on simple feedback-driven approaches, which have proven inadequate in practical settings. \skel~\cite{wang2025skel}, \oxidizer~\cite{zhang2025oxidizer}, and \syzygy~\cite{shetty2024syzygy} utilize multi-turn iterative prompting techniques. \alphatrans~\cite{ibrahimzada2025alphatrans} adopts an execution trace-based reprompting strategy. 
For instance, consider scenario~(\circled[fill=red]{c1}), where running the \texttt{\small test} method sequentially invokes methods \texttt{\small m1}, \texttt{\small m2}, and \texttt{\small m3}. If a bug is present in fragment \texttt{\small m2}, \alphatrans reprompts all executed fragments individually without considering inter-fragment dependencies. This approach, while potentially resolving the bug in \texttt{\small m2}, risks introducing new functional bugs in previously correct fragments, such as \texttt{\small m1}. To overcome this limitation, \approach uses code analysis and an LLM agent ~(\circled[fill=red]{c2}).
The analyses expose dependencies such as the one between \texttt{\small m1} and \texttt{\small m2} and the agent interacts with the execution environment to execute, validate, and iteratively refine its generated patches.

\section{Experimental Setup}
\label{appendix:experiment-setup}

\begin{table}[t]
    \centering
    \caption{Details of benchmarks\protect\tablefootnote{Some projects in \skel\ (e.g., \texttt{colorsys}) are part of a bigger project~\cite{cpython}. The reported Star and Fork numbers belong to that bigger project.} from existing techniques used in \approach. \textbf{LoC:} Lines of code in the source project.}
    \resizebox{\textwidth}{!}{
        \begin{tabular}{llccccccc} 
\hline
\textbf{Tool}                                                                 & \textbf{Project}                     & \begin{tabular}[c]{@{}c@{}}\textbf{Source}\\\textbf{Language }\end{tabular} & \begin{tabular}[c]{@{}c@{}}\textbf{Target}\\\textbf{Language }\end{tabular} & \begin{tabular}[c]{@{}c@{}}\textbf{Total \#}\\\textbf{Trans. Pairs}\end{tabular} & \textbf{LoC}    & \begin{tabular}[c]{@{}c@{}}\textbf{Test}\\\textbf{Coverage (\%)}\end{tabular} & \textbf{Stars}  & \textbf{Forks}   \\ 
\hline
\multirow{6}{*}{\oxidizer~\cite{zhang2025oxidizer}}           & checkdigit~\cite{checkdigit}         & \multirow{6}{*}{Go}                                                         & \multirow{6}{*}{Rust}                                                       & 29                                                                               & 428             & 86.2                                                                          & 111             & 8                \\
                                                                              & go-edlib~\cite{goedlib}              &                                                                             &                                                                             & 24                                                                               & 639             & 100                                                                           & 517             & 27               \\
                                                                              & histogram~\cite{gohistogram}         &                                                                             &                                                                             & 19                                                                               & 314             & 68.4                                                                          & 176             & 31               \\
                                                                              & nameparts~\cite{gonameparts}         &                                                                             &                                                                             & 15                                                                               & 413             & 100                                                                           & 43              & 5                \\
                                                                              & stats~\cite{stats}                   &                                                                             &                                                                             & 53                                                                               & 1241            & 98.1                                                                          & 2989            & 170              \\
                                                                              & textrank~\cite{textrank}             &                                                                             &                                                                             & 52                                                                               & 1132            & 100                                                                           & 217             & 22               \\ 
\hline
\multirow{4}{*}{\alphatrans~\cite{ibrahimzada2025alphatrans}} & cli~\cite{commons-cli}               & \multirow{4}{*}{Java}                                                       & \multirow{4}{*}{Python}                                                     & 273                                                                              & 37841           & 93.8                                                                          & 372             & 201              \\
                                                                              & csv~\cite{commons-csv}               &                                                                             &                                                                             & 235                                                                              & 33072           & 88.1                                                                          & 392             & 278              \\
                                                                              & fileupload~\cite{commons-fileupload} &                                                                             &                                                                             & 192                                                                              & 3567            & 20.8                                                                          & 246             & 185              \\
                                                                              & validator~\cite{commons-validator}   &                                                                             &                                                                             & 646                                                                              & 41605           & 62.5                                                                          & 216             & 164              \\ 
\hline
\multirow{8}{*}{\skel~\cite{wang2025skel}}                    & bst~\cite{bst}                       & \multirow{8}{*}{Python}                                                     & \multirow{8}{*}{JavaScript}                                                 & 19                                                                               & 123             & 100                                                                           & 203000          & 47000            \\
                                                                              & colorsys~\cite{colorsys}             &                                                                             &                                                                             & 8                                                                                & 120             & 100                                                                           & 67900           & 32300            \\
                                                                              & heapq~\cite{heapq}                   &                                                                             &                                                                             & 22                                                                               & 189             & 100                                                                           & 67900           & 32300            \\
                                                                              & html~\cite{html}                     &                                                                             &                                                                             & 44                                                                               & 684             & 95.5                                                                          & 67900           & 32300            \\
                                                                              & mathgen~\cite{mathgen}               &                                                                             &                                                                             & 81                                                                               & 735             & 100                                                                           & 711             & 183              \\
                                                                              & rbt~\cite{rbt}                       &                                                                             &                                                                             & 27                                                                               & 366             & 96.3                                                                          & 203000          & 47000            \\
                                                                              & strsim~\cite{strsim}                 &                                                                             &                                                                             & 64                                                                               & 654             & 78.1                                                                          & 1014            & 125              \\
                                                                              & toml~\cite{toml}                     &                                                                             &                                                                             & 72                                                                               & 1206            & 65.3                                                                          & 1126            & 192              \\ 
\hline
\multirow{6}{*}{\rustrepotrans~\cite{ou2024repository}}       & charset~\cite{charset-normalizer}    & Python                                                                      & \multirow{6}{*}{Rust}                                                       & 33                                                                               & 4231            & 100                                                                           & 672             & 56               \\
                                                                              & deltachat~\cite{deltachat}           & C                                                                           &                                                                             & 125                                                                              & 23116           & 98.4                                                                          & 306             & 28               \\
                                                                              & iceberg-java \cite{iceberg-java}     & Java                                                                        &                                                                             & 25                                                                               & 592793          & 100                                                                           & 7700            & 2700             \\
                                                                              & iceberg-py \cite{iceberg-python}     & Python                                                                      &                                                                             & 44                                                                               & 49746           & 97.7                                                                          & 805             & 327              \\
                                                                              & crypto-c \cite{crypto-c}             & C                                                                           &                                                                             & 20                                                                               & 5922            & 100                                                                           & 36              & 15               \\
                                                                              & crypto-java \cite{crypto-java}       & Java                                                                        &                                                                             & 97                                                                               & 110261          & 100                                                                           & 2               & 7                \\ 
\hline
\textbf{Total}                                                                &                                      &                                                                             &                                                                             & \textbf{2219}                                                                    & \textbf{910398} & \textbf{89.6}                                                                 & \textbf{627351} & \textbf{195624}  \\
\hline
\end{tabular}
    }
    \label{table:stats}
\end{table}

\subsection{Benchmark}
We evaluate \approach on benchmarks used in prior work on automated repository-level translation. Each benchmark problem is a translation pair: the source function and the corresponding translation. The task for each benchmark problem is to give a verdict on the functional equivalence of pairs in two different PLs, and repair translation in case of equivalence. Our subjects are open-source repository-level translations with equivalence verdicts available\footnote{We exclude rule-based transpilers as they do not include a validation mechanism, i.e., their translation is (theoretically) correct by construction.} from the peer-reviewed literature\footnote{This criterion excludes techniques such as RustMap~\cite{cai2025rustmap} and C2SaferRust~\cite{nitin2026c2saferrust}.}. We make selections across a diverse set of PL pairs.


Table~\ref{table:stats} summarizes our subject translation pairs. We collect subjects from three recent repository-level code translation techniques~\cite{ibrahimzada2025alphatrans,zhang2025oxidizer,wang2025skel}. We did not include \syzygy~\cite{shetty2024syzygy} because its artifact does not provide a validation system for individual functions 
(it only provides end-to-end tests). 
These works generated translations of real-world open source GitHub projects, and performed equivalence validation at the individual function level.
Since \alphatrans~\cite{ibrahimzada2025alphatrans} is evaluated on a large number of functions, we randomly sample $1346$ of its total $4643$ translation pairs\footnote{\alphatrans translated both application and test code. In this work, we only included the application code translation pairs. While reviewing their artifacts, we also noted $11$ translation pairs to be dead code and excluded them.}. 

To demonstrate the adaptability of \approach to more PL pairs, we also collect subjects from \rustrepotrans~\cite{ou2024repository}, a benchmark consisting of human-written translations into Rust and 
unit tests. 
We exclude translation pairs collected from the libp2p~\cite{libp2p} projects in \rustrepotrans due to the presence of non-deterministic flaky 
tests that may result in false negatives, i.e., functional inequivalence while translation is correct, unfairly biasing comparison in favor of \approach. In total, we collect $2{,}219$ translation pairs with over $900K$ lines of code from $24$ projects and in $6$ different PL pairs.

\subsection{LLMs} 
Major software engineering leaderboards~\cite{swebench,bigcodebench} have shown that Claude Sonnet~\cite{claude} outperforms
other 
proprietary LLMs, such as OpenAI GPT-4o~\cite{gpt4o} and Google Gemini Pro~\cite{gemini}. Therefore, we use Anthropic's \texttt{\small Claude 3.7 Sonnet}~\cite{claude} and Claude Code ($1.0.51$)~\cite{claudecode} as the main LLM and agent in all our experiments. To show the adaptability of \approach to different LLMs and agentic frameworks, we repeat a subset of experiments using OpenAI \texttt{\small o4-mini-2025-04-16}~\cite{o4mini} and Codex~\cite{codex} (\S\ref{subsec:rq3}). 
To support future research and external validation, \approach logs the inputs, intermediate agent interactions, tool execution results, and outputs of the LLM, and supports visualizing and inspecting these logs. \approach terminates within the budget of $1{,}000$ seconds. We empirically set this timeout after analyzing the execution time of $300$ samples.

\subsection{Competing Validation \& Repair Tools}
We compare \approach's validation technique with the automated validation techniques proposed by \skel~\cite{wang2025skel}, \oxidizer~\cite{zhang2025oxidizer}, and \alphatrans~\cite{ibrahimzada2025alphatrans}. Except \rustrepotrans, other approaches do not explicitly report the repair results, as the repair process is interleaved with translation in the loop. For those techniques~\cite{ibrahimzada2025alphatrans, zhang2025oxidizer, wang2025skel}, we compare \approach repair results with their final translation success.


\subsection{\approach Implementation} We implement our structure-based semantic 
analysis on top of Tree-Sitter~\cite{tree-sitter}, as it supports $165+$ languages, including six PLs we target in this study. 
For running tests and validating patches, \approach uses Rust $1.87.0$~\cite{rustlang}, Python $3.12.9$~\cite{pythonlang}, Java $21.0.7$~\cite{javalang}, Node $22.16.0$~\cite{nodejs}, GCC $7.3.1$~\cite{gcc}, and Go $1.24.4$~\cite{golang}. 

\section{Analysis of (Dis)Agreements and Examples of Incorrect Verdicts for Claude Experiments}
\label{sec:claude-analysis-disagreements-examples}

\subsection{Analysis of the Disagreements}
\label{subsec:rq1-dispute-analysis}

Both \approach and existing validation approaches are prone to false positives (i.e. the verdict is equivalent but the translation is not) and false negatives. To determine false positives and false negatives, 
we perform a manual investigation.

We first categorize disagreements into two cases: $D_1$, where \approach produced a \emph{not equivalent} verdict and the other disagreed; and $D_2$, where \approach produced an \emph{equivalent} verdict and the other disagreed. Due to large number of studied translation pairs, we randomly sample five instances of both $D_1$ and $D_2$ from each of the $24$ projects. If a project had fewer than five disagreements, we considered all instances without sampling. Two authors\footnote{The selected authors have total experience of $17$ years in academia and $4.5$ years in industry.} independently reviewed disagreements to verify whether \emph{the not equivalent verdict} was correct (by \approach in $D_1$ and by others in $D_2$) and achieved an inter-rater agreement of $83.7\%$. If the not equivalent verdict is correct, the reviewer rules it in favor of the tool that said not equivalent, otherwise they rule it favor of the tool that said equivalent. During the investigation, $3$ disagreements with \oxidizer were due to the tool's function mocking not being enabled, and were unable to enable it. In addition, $11$ disagreements with \rustrepotrans were due to the correct translation not being \texttt{\small ``1:1''}, or in other words, the correct translation is not functionally equivalent to the source function. These disagreements would have been unfairly ruled in favor of \approach, and so were filtered out, leaving us with $145$ disagreement cases ($D_1=92$, $D_2=53$). When the reviewer's resolution conflicted ($18.6\%$ of cases), they met with each other and agreed on the final resolution\footnote{The results of our human investigation with comments by each reviewer are publicly available~\cite{website}.}.

The \textbf{Disagreement} columns in Table~\ref{table:rq1-validation} summarize the result of our human investigation. The \textbf{Tool} column shows the percentage of disagreements ruled in favor of the existing validation tool, and the \textbf{Ours} columns shows the percentage ruled in favor of \approach. The disagreement resolutions show that \approach’s verdicts are often more accurate than existing automated validation tools. \approach's verdicts are significantly more accurate than \oxidizer and \alphatrans --- disagreements are ruled in favor of \approach in $84.1\%$ and $73.5\%$ of cases, respectively. Compared to \skel, \approach shows similar accuracy ($53.5\%$). On \rustrepotrans, \approach's accuracy fares worse ($12.5\%$), which is due to the relative complexity of its translation pairs. 

Existing validation approaches produced $88$ incorrect verdicts, $80$ of which fell into three categories: 
(1) \textbf{Inadequate Unit Tests ($\textbf{42}$ of cases)},
(2) \textbf{Excessively Strict Equivalence Definition ($\textbf{11}$ of cases)}, 
and (3) \textbf{Language Interoperability Bug ($\textbf{27}$ of cases)}.
A language interoperability bug means that the tool's process for converting concrete inputs in the source language to the target language did not preserve equivalence. For example, \oxidizer's conversion of \texttt{\small Go} \texttt{\small rune}s (which represent a Unicode character) to \texttt{\small Rust} \texttt{\small char}s resulted in different Unicode characters. 

On the benchmarks associated with automated validation tools (\skel, \oxidizer, \alphatrans), \approach produced 36 incorrect verdicts, $34$ of which fell into three categories: 
(1) \textbf{Hallucination ($\textbf{23}$ cases)}, 
(2) \textbf{Inadequate Unit Tests ($\textbf{4}$ cases)} (meaning \approach missed an input that would demonstrate inequivalence), 
(3) \textbf{Infeasible Input ($\textbf{7}$ cases)}.
An infeasible input means that the LLM discovered an input where the source function and translation produce different outputs, but the input can never occur when the project is used as intended. Infeasible inputs often involve directly initializing private class/struct members, or calling private helper methods in unintended ways.

On \rustrepotrans's benchmarks, \approach produced $21$ incorrect verdicts. Two of the main causes are similar to the other benchmarks: (1) \textbf{Hallucination ($\textbf{7}$ cases)} and (2) \textbf{Inadequate Unit Tests ($\textbf{6}$ cases)}. The increased rates of these two causes are due to the relative complexity and size of \rustrepotrans's projects. The other major cause is \textbf{Excessively Strict Equivalence Definition ($\textbf{6}$ cases)}. As previously mentioned, \rustrepotrans's translations include refactors to make the translation more idiomatic, which creates ambiguity around the proper definition equivalence.


\subsection{Representative Examples of Incorrect Verdicts}
\label{subsec:rq1-incorrect-verdicts}

The following code snippet shows an example of \textbf{Inadequate Unit Tests} on a translation pair from \oxidizer. The functions both calculate the edit distance between two strings. The functions are not equivalent because \texttt{\small Go}'s \texttt{\small len()} function counts Unicode characters, whereas \texttt{\small Rust}'s \texttt{\small .len()} function counts bytes. \oxidizer incorrectly validates the \texttt{\small Rust} translation as equivalent because the source project's unit tests do not cover non-ASCII inputs. However, \approach successfully generates a test with non-ASCII inputs demonstrating they are not equivalent.

\noindent
\begin{minipage}{.475\linewidth}
\begin{minted}[frame=lines,framesep=1mm,baselinestretch=0.5, fontsize=\scriptsize, breaklines, breakanywhere, linenos,numbersep=2pt]{go}
------------------- GO SOURCE CODE -------------------
func LCSEditDistance(str1 string, str2 string)      int {
    
    // ... if conditions ...
    lcs := LCS(s1, s2)
    return (len([]rune(s1)) - lcs) +          (len([]rune(s2)) - lcs)
}
\end{minted}
\end{minipage}\hfill
\begin{minipage}{.475\linewidth}
\begin{minted}[escapeinside=||, frame=lines,framesep=1mm,baselinestretch=0.5, fontsize=\scriptsize, breaklines, breakanywhere, linenos,numbersep=2pt]{rust}
------------------ RUST TRANSLATION ------------------
pub fn lcs_edit_distance(str1: &str, str2: &str) -> Result<i32> {
    // ... if conditions ...
    let lcs_len = lcs(str1, str2)?;
    let edit_distance = (str1.len() as i32 - lcs_len) + (str2.len() as i32 - lcs_len);
    Ok(edit_distance)
}
\end{minted}
\end{minipage}

The next code snippet demonstrates an example of an \textbf{Infeasible Input} taken from the \texttt{\small textrank} project.
The below \texttt{\small Go} function inserts a value into the map \texttt{\small ranks.SentenceMap} (respectively, \texttt{\small ranks.sentence\_map} for the \texttt{\small Rust} translation). 
\approach discovers that these functions return different values when the map is initially \texttt{\small \{ 1 : ``SomeString'' \}} (the \texttt{\small Go} returns 0 while the \texttt{\small Rust} returns 1). 
However, when the \texttt{\small textrank} is used properly via its public interface, this initial state for the map cannot occur. 
The map will always contain keys from $1$ to $n$, where $n$ is the number of map entries. Under this precondition, the functions are equivalent.

\noindent
\begin{minipage}{.475\linewidth}
\begin{minted}[frame=lines,framesep=1mm,baselinestretch=0.5, fontsize=\scriptsize, breaklines, breakanywhere, linenos,numbersep=2pt]{go}
------------------- GO SOURCE CODE -------------------
func addSentence(ranks *Rank, sentence ParsedSentence) int {,
    ranks.SentenceMap[len(ranks.SentenceMap)] = sentence.GetOriginal()
    return len(ranks.SentenceMap) - 1

}
\end{minted}
\end{minipage}\hfill
\begin{minipage}{.475\linewidth}
\begin{minted}[escapeinside=||, frame=lines,framesep=1mm,baselinestretch=0.5, fontsize=\scriptsize, breaklines, breakanywhere, linenos,numbersep=2pt]{rust}
------------------ RUST TRANSLATION ------------------
pub(crate) fn add_sentence(ranks: &mut Rank, sentence: ParsedSentence) -> Result<i32, Error> {
    let sentence_id = ranks.sentence_map.len() as i32;
    ranks.sentence_map.insert(sentence_id, sentence.original.clone());
    Ok(sentence_id)
}
\end{minted}
\end{minipage}

The next code snippet demonstrates an example of an \textbf{Excessively Strict Equivalence Definition} taken from the \texttt{\small deltachat-core} project. Both the \texttt{\small C} and \texttt{\small Rust} function retrieve a field \texttt{\small blobdir}. \approach states that these are not equivalent because (1) the \texttt{\small Rust} function does not perform a null check, and (2) the \texttt{\small C} function returns a copy of \texttt{\small blobdir} whereas the \texttt{\small Rust} returns a reference. While this is true, the translation follows \texttt{\small Rust}'s idioms, and a developer would not care about these differences. \texttt{\small Rust}'s type system prevents \texttt{\small blobdir} from ever being null, and the \texttt{\small Rust} translation returns an \textit{immutable} reference, preventing the caller from modifying the return value. 

\noindent
\begin{minipage}{.475\linewidth}
\begin{minted}[frame=lines,framesep=1mm,baselinestretch=0.5, fontsize=\scriptsize, breaklines, breakanywhere, linenos,numbersep=2pt]{c}
------------------- C SOURCE CODE --------------------
char* dc_get_blobdir(const dc_context_t* context) {
    if (context==NULL || context->magic!=DC_CONTEXT_MAGIC) {
        return dc_strdup(NULL);
    }
    return dc_strdup(context->blobdir);
}
\end{minted}
\end{minipage}\hfill
\begin{minipage}{.475\linewidth}
\begin{minted}[escapeinside=||, frame=lines,framesep=1mm,baselinestretch=0.5, fontsize=\scriptsize, breaklines, breakanywhere, linenos,numbersep=2pt]{rust}
------------------ RUST TRANSLATION ------------------
pub fn get_blobdir(&self) -> &Path {
    
    
    &self.inner.blobdir



}
\end{minted}
\end{minipage}

\subsection{Analysis of the Agreements}
\label{subsec:rq1-non-dispute-analysis}

From the original experiments using \texttt{\small Claude 3.7 Sonnet}, we uniformly sampled at most $5$ equivalent validations across $24$ benchmarks and four validation systems, resulting in $110$ source--target pairs. In total, $88$ were correctly validated as equivalent, and $22$ validated as non-equivalent by both existing tools and \approach. The same two authors from dispute analysis (\S\ref{subsec:rq1-dispute-analysis}) performed manual investigation of equivalent cases with $95.5\%$ inter-rater agreement and show that $\frac{106}{110}$ ($96.4\%$) judgements were correct, meaning $\frac{4}{110}$ ($3.6\%$) were incorrect validations, indicating a need for more rigorous testing and oversight in \approach.

The following code snippets show an example of false positive, where both \approach and \oxidizer validate the translation as equivalent, but the human investigator concludes otherwise. The translated Rust code incorrectly returns an error if any word is not found in \texttt{\small word\_val\_id}, while the Go source code silently handles missing words and returns zero for missing keys and continues. This difference in error handling means they behave differently when words are not in the map. \oxidizer and \approach failed to detect this bug due to insufficient testing.

\noindent
\begin{minipage}{.475\linewidth}
\begin{minted}[frame=lines,framesep=1mm,baselinestretch=0.5, fontsize=\scriptsize, breaklines, breakanywhere, linenos,numbersep=2pt]{go}
------------------- GO SOURCE CODE -------------------
func FindSentencesByPhrases(ranks *Rank, words []string) []Sentence {




    for _, i := range words {
        for _, j := range words {
            x := ranks.WordValID[i]
            y := ranks.WordValID[j]
            ...
        }
    }



    ...
    return sentences
}
\end{minted}
\end{minipage}\hfill
\begin{minipage}{.475\linewidth}
\begin{minted}[escapeinside=||, frame=lines,framesep=1mm,baselinestretch=0.5, fontsize=\scriptsize, breaklines, breakanywhere, linenos,numbersep=2pt]{rust}
------------------ RUST TRANSLATION ------------------
pub fn find_sentences_by_phrases(ranks: Option<&Rank>, words: &[String]) -> Result<Vec<Sentence>> {

    for i in words {
        for j in words {
            let x = *ranks.word_val_id.get(i)
            .ok_or_else(|| anyhow::
            anyhow!("Word not found: {}", i))?;
            let y = *ranks.word_val_id.get(j)
            .ok_or_else(|| anyhow::
            anyhow!("Word not found: {}", j))?;
            ...
        }
    }
    ...
    Ok(sentences)
}
\end{minted}
\end{minipage}

The next code snippets demonstrate an example of false negative, where both \approach and \rustrepotrans validate the translation as non-equivalent, but the human investigator concludes otherwise. Both functions compute the current time in slots since the Unix epoch by dividing the current Unix timestamp (in seconds) by \texttt{\small (60 * TIME\_SLOT\_MINUTES)}. The logic is identical, with only language-specific differences in how current time is obtained and types used (\texttt{\small unsign32} vs \texttt{\small usize}). \rustrepotrans and \approach validate these as functionally inequivalent because there is a compilation bug somewhere else in the codebase.

\noindent
\begin{minipage}{.475\linewidth}
\begin{minted}[frame=lines,framesep=1mm,baselinestretch=0.5, fontsize=\scriptsize, breaklines, breakanywhere, linenos,numbersep=2pt]{c}
------------------- C SOURCE CODE --------------------
    unsign32 today(void)
    {
        /* return time in slots since epoch */
        unsign32 ti = (unsign32) time(NULL);
        return ti / (60 * TIME_SLOT_MINUTES);
    }
\end{minted}
\end{minipage}\hfill
\begin{minipage}{.475\linewidth}
\begin{minted}[escapeinside=||, frame=lines,framesep=1mm,baselinestretch=0.5, fontsize=\scriptsize, breaklines, breakanywhere, linenos,numbersep=2pt]{rust}
------------------ RUST TRANSLATION ------------------
    pub fn today() -> usize {
        // Return time in slots since epoch
        let now = SystemTime::now().duration_since(UNIX_EPOCH).unwrap().as_secs() as usize;
        now / (60 * rom::TIME_SLOT_MINUTES)
    }
\end{minted}
\end{minipage}

\section{Threshold Ablation}
\label{sec:threshold-ablation}

Table~\ref{table:threshold-ablation} shows the results of our ablation study on the threshold $\tau$ in Algorithm~\ref{alg:cfganalyzer} and Algorithm~\ref{alg:dfpathanalyzer}. The threshold $\tau$ governs a tradeoff between algorithmic reliance and LLM invocation cost in the Control Flow and Data Flow Path analyzers, which are the only two sub-analyzers with a short-circuit mechanism; the remaining four (I/O Mapping, Library API, Exception Handling, Specifications) are unaffected and serve as an internal control. A lower $\tau$ is easier to satisfy, causing the short-circuit to fire more frequently and reducing LLM cost, but over-trusting the structural similarity algorithm: at $\tau{=}0.5$, TNR on Control Flow and Data Flow Path drops to $19.7\%$ and $17.7\%$ respectively---meaning the structural algorithm alone fails to correctly identify inequivalent pairs in roughly four out of five cases. A higher $\tau$ is harder to satisfy, so fewer cases are short-circuited and more LLM invocations are triggered, increasing cost. Moving from $\tau{=}0.5$ to $\tau{=}0.7$ yields the largest correctness gains: TNR on Control Flow jumps $+14.6\%$ at a TPR cost of only $-2.6\%$, and TNR on Data Flow Path gains $+5.6\%$ at a negligible $-0.2\%$ TPR cost. Moving further to $\tau{=}0.9$ yields only marginal additional TNR improvements ($+8.4\%$ and $+3.5\%$ respectively) while TPR continues to erode ($-1.6\%$ and $-1.4\%$) and LLM invocation cost increases. The average TNR across all six analyzers follows the same pattern: $34.3\%$ at $\tau{=}0.5$, $37.8\%$ at $\tau{=}0.7$, and $39.8\%$ at $\tau{=}0.9$, confirming that $\tau{=}0.7$ captures the steepest part of the improvement curve. We therefore select $\tau{=}0.7$ as the main threshold value---delivering substantial TNR improvement over $\tau{=}0.5$ without the diminishing returns and added cost of $\tau{=}0.9$. Nonetheless, since the threshold functions as a cost-optimization lever rather than a correctness lever---with any errors in a single sub-analyzer recoverable by the downstream Test Generator \& Repair Agent and Verdict Agent---practitioners can adjust $\tau$ based on their own risk tolerance and computational budget, using the TPR/TNR/FPR/FNR results in Table~\ref{table:threshold-ablation}.

\begin{table}[!htbp]
    \centering
    \caption{Agreement between each semantic analyzer and \approach verdict. Experiments repeated with the Claude model for threshold $\tau$ $\in$ $\{0.5, 0.7, 0.9\}$. \textbf{TP:} True Positive (Semantic Analyzer=Yes and \approach=Yes), \textbf{FP:} False Positive (Semantic Analyzer=Yes and \approach=No), \textbf{FN:} False Negative (Semantic Analyzer=No and \approach=Yes), \textbf{TN:} True Negative (Semantic Analyzer=No and \approach=No).}
    \begin{tabular}{c|c|cc|cc|cc|cc}
\hline
\multirow{2}{*}{\textbf{Semantic Analyzer}} & \multirow{2}{*}{\textbf{$\boldsymbol{\tau}$}} & \multicolumn{2}{c|}{\textbf{TP}} & \multicolumn{2}{c|}{\textbf{FP}} & \multicolumn{2}{c|}{\textbf{FN}} & \multicolumn{2}{c}{\textbf{TN}} \\
\cline{3-10}
 & & \multicolumn{1}{c|}{\textbf{Count}} & \textbf{TPR} & \multicolumn{1}{c|}{\textbf{Count}} & \textbf{FPR} & \multicolumn{1}{c|}{\textbf{Count}} & \textbf{FNR} & \multicolumn{1}{c|}{\textbf{Count}} & \textbf{TNR} \\
\hline
\multirow{3}{*}{Control Flow}             & 0.5 & 1396 & 0.957 & 549 & 0.803 & 63  & 0.043 & 135 & 0.197 \\
                                          & 0.7 & 1404 & 0.931 & 438 & 0.657 & 104 & 0.069 & 229 & 0.343 \\
                                          & 0.9 & 1342 & 0.915 & 375 & 0.573 & 124 & 0.085 & 279 & 0.427 \\
\hline
\multirow{3}{*}{Data Flow Path}           & 0.5 & 1411 & 0.968 & 552 & 0.823 & 47  & 0.032 & 119 & 0.177 \\
                                          & 0.7 & 1459 & 0.966 & 510 & 0.767 & 52  & 0.034 & 155 & 0.233 \\
                                          & 0.9 & 1395 & 0.952 & 483 & 0.732 & 70  & 0.048 & 177 & 0.268 \\
\hline
\multirow{3}{*}{Input/Output Mapping}     & 0.5 & 1399 & 0.961 & 323 & 0.474 & 57  & 0.039 & 358 & 0.526 \\
                                          & 0.7 & 1437 & 0.953 & 321 & 0.481 & 71  & 0.047 & 347 & 0.519 \\
                                          & 0.9 & 1416 & 0.963 & 322 & 0.481 & 55  & 0.037 & 348 & 0.519 \\
\hline
\multirow{3}{*}{Library API Equivalence}  & 0.5 & 1441 & 0.992 & 478 & 0.714 & 12  & 0.008 & 191 & 0.286 \\
                                          & 0.7 & 1481 & 0.988 & 482 & 0.730 & 18  & 0.012 & 178 & 0.270 \\
                                          & 0.9 & 1450 & 0.989 & 477 & 0.726 & 16  & 0.011 & 180 & 0.274 \\
\hline
\multirow{3}{*}{Exception/Error Handling} & 0.5 & 1396 & 0.960 & 433 & 0.640 & 58  & 0.040 & 244 & 0.360 \\
                                          & 0.7 & 1420 & 0.940 & 425 & 0.636 & 91  & 0.060 & 243 & 0.364 \\
                                          & 0.9 & 1407 & 0.959 & 428 & 0.641 & 60  & 0.041 & 240 & 0.359 \\
\hline
\multirow{3}{*}{Specifications}           & 0.5 & 1378 & 0.953 & 329 & 0.486 & 68  & 0.047 & 348 & 0.514 \\
                                          & 0.7 & 1406 & 0.938 & 308 & 0.462 & 93  & 0.062 & 359 & 0.538 \\
                                          & 0.9 & 1376 & 0.947 & 308 & 0.460 & 77  & 0.053 & 362 & 0.540 \\
\hline\hline
\multirow{3}{*}{Average}                  & 0.5 & 1403 & 0.965 & 444 & 0.657 & 50  & 0.035 & 232 & 0.343 \\
                                          & 0.7 & 1434 & 0.953 & 414 & 0.622 & 71  & 0.047 & 251 & 0.378 \\
                                          & 0.9 & 1397 & 0.954 & 398 & 0.602 & 67  & 0.046 & 264 & 0.398 \\
\hline
\end{tabular}    
    \label{table:threshold-ablation}
\end{table}

\section{\approach with Open Source Model}
\label{sec:open-model-experiments}

\approach is LLM-agnostic and easily adapts to new models. To show its open-source adaptability, we reproduced Table~\ref{table:rq1-validation} with the state-of-the-art \texttt{\small Qwen3-Next-80B-A3B} (released February 2026). Table~\ref{table:rq1-validation-qwen} shows the results of this experiment, indicating that the agreement rate of Qwen3 ($71.6\%$) remains consistent with Claude ($72.8\%$), suggesting similarity between the open-source and closed LLMs.

\begin{table}[!htbp]
    \scriptsize
    \centering
    \caption{Effectiveness of \approach in translation validation compared to existing techniques using \texttt{\small Qwen3-Next-80B-A3B}. \textbf{EQ:}~Equivalent, \textbf{NEQ:} Not Equivalent, \textbf{VF:} Validation Failure, \textbf{Agreement}: number and percentage of translation pairs where \approach's and the existing tool's verdicts agree, \textbf{Disagreement}: percentage of disagreements ruled in favor of \textbf{Tool} and \approach (\textbf{Ours}). \textbf{VF}s are excluded from \textbf{Agreement} and \textbf{Disagreement} calculations.}
    \begin{tabular}{c|lc|ccc|ccc|c} 
\hline
\multicolumn{1}{l|}{\multirow{2}{*}{\textbf{Tool}}}                                                         & \multirow{2}{*}{\textbf{Project}} & \multirow{2}{*}{\begin{tabular}[c]{@{}c@{}}\textbf{Total \#}\\\textbf{Trans. Pairs}\end{tabular}} & \multicolumn{3}{c|}{\textbf{Tool Validation }}                                             & \multicolumn{3}{c|}{\textbf{\approach}}                                                  & \multirow{2}{*}{\textbf{Agreement}}  \\ 
\cline{4-9}
\multicolumn{1}{l|}{}                                                                                       &                                   &                                                                                                   & \multicolumn{1}{c|}{\textbf{EQ}} & \multicolumn{1}{c|}{\textbf{NEQ}} & \textbf{VF}         & \multicolumn{1}{c|}{\textbf{EQ}} & \multicolumn{1}{c|}{\textbf{NEQ}} & \textbf{VF}       &                                      \\ 
\hline
\multirow{6}{*}{\rotatebox{90}{\oxidizer}}                                                                  & checkdigit                        & 29                                                                                                & 21 (72.4)                        & 8 (27.6)                          & 0 (0)               & 23 (79.3)                        & 6 (20.7)                          & 0 (0)             & 19 (65.5)                            \\
                                                                                                            & go-edlib                          & 24                                                                                                & 18 (75)                          & 6 (25)                            & 0 (0)               & 14 (58.3)                        & 10 (41.7)                         & 0 (0)             & 12 (50)                              \\
                                                                                                            & histogram                         & 19                                                                                                & 12 (63.2)                        & 7 (36.8)                          & 0 (0)               & 14 (73.7)                        & 5 (26.3)                          & 0 (0)             & 7 (36.8)                             \\
                                                                                                            & nameparts                         & 15                                                                                                & 9 (60)                           & 6 (40)                            & 0 (0)               & 10 (66.7)                        & 5 (33.3)                          & 0 (0)             & 14 (93.3)                            \\
                                                                                                            & stats                             & 53                                                                                                & 38 (71.7)                        & 14 (26.4)                         & 1 (1.9)             & 35 (66)                          & 17 (32.1)                         & 1 (1.9)           & 33 (64.7)                            \\
                                                                                                            & textrank                          & 52                                                                                                & 40 (76.9)                        & 12 (23.1)                         & 0 (0)               & 32 (61.5)                        & 20 (38.5)                         & 0 (0)             & 32 (61.5)                            \\ 
\hline
\rowcolor[rgb]{0.8,0.902,0.902} \multicolumn{1}{l|}{\textbf{Total}}                                         &                                   & \textbf{192}                                                                                      & \textbf{138 (71.9)}              & \textbf{53 (27.6)}                & \textbf{1 (0.5)}    & \textbf{128 (66.7)}              & \textbf{63 (32.8)}                & \textbf{1 (0.5)}  & \textbf{117 (61.6)}                  \\ 
\hline
\multirow{4}{*}{\rotatebox{90}{\begin{tabular}[c]{@{}c@{}}\textsc{Alpha}\\ \textsc{Trans}\end{tabular}}}    & cli                               & 273                                                                                               & 210 (76.9)                       & 24 (8.8)                          & 39 (14.3)           & 214 (78.4)                       & 53 (19.4)                         & 6 (2.2)           & 173 (75.2)                           \\
                                                                                                            & csv                               & 235                                                                                               & 97 (41.3)                        & 61 (26)                           & 77 (32.8)           & 155 (66)                         & 75 (31.9)                         & 5 (2.1)           & 93 (60)                              \\
                                                                                                            & fileupload                        & 192                                                                                               & 19 (9.9)                         & 1 (0.5)                           & 172 (89.6)          & 143 (74.5)                       & 43 (22.4)                         & 6 (3.1)           & 17 (85)                              \\
                                                                                                            & validator                         & 646                                                                                               & 247 (38.2)                       & 103 (15.9)                        & 296 (45.8)          & 475 (73.5)                       & 158 (24.5)                        & 13 (2)            & 223 (65)                             \\ 
\hline
\rowcolor[rgb]{0.8,0.902,0.902} \multicolumn{1}{l|}{\textbf{Total}}                                         &                                   & \textbf{1346}                                                                                     & \textbf{573 (42.6)}              & \textbf{189 (14)}                 & \textbf{584 (43.4)} & \textbf{987 (73.3)}              & \textbf{329 (24.4)}               & \textbf{30 (2.2)} & \textbf{506 (67.6)}                  \\ 
\hline
\multirow{8}{*}{\rotatebox{90}{\skel}}                                                                      & bst                               & 19                                                                                                & 19 (100)                         & 0 (0)                             & 0 (0)               & 13 (68.4)                        & 5 (26.3)                          & 1 (5.3)           & 13 (72.2)                            \\
                                                                                                            & colorsys                          & 8                                                                                                 & 8 (100)                          & 0 (0)                             & 0 (0)               & 8 (100)                          & 0 (0)                             & 0 (0)             & 8 (100)                              \\
                                                                                                            & heapq                             & 22                                                                                                & 19 (86.4)                        & 3 (13.6)                          & 0 (0)               & 19 (86.4)                        & 3 (13.6)                          & 0 (0)             & 20 (90.9)                            \\
                                                                                                            & html                              & 44                                                                                                & 40 (90.9)                        & 2 (4.5)                           & 2 (4.5)             & 23 (52.3)                        & 19 (43.2)                         & 2 (4.5)           & 24 (60)                              \\
                                                                                                            & mathgen                           & 81                                                                                                & 77 (95.1)                        & 4 (4.9)                           & 0 (0)               & 66 (81.5)                        & 12 (14.8)                         & 3 (3.7)           & 66 (84.6)                            \\
                                                                                                            & rbt                               & 27                                                                                                & 26 (96.3)                        & 0 (0)                             & 1 (3.7)             & 19 (70.4)                        & 5 (18.5)                          & 3 (11.1)          & 18 (78.3)                            \\
                                                                                                            & strsim                            & 64                                                                                                & 50 (78.1)                        & 0 (0)                             & 14 (21.9)           & 56 (87.5)                        & 6 (9.4)                           & 2 (3.1)           & 45 (91.8)                            \\
                                                                                                            & toml                              & 72                                                                                                & 37 (51.4)                        & 10 (13.9)                         & 25 (34.7)           & 46 (63.9)                        & 23 (31.9)                         & 3 (4.2)           & 35 (77.8)                            \\ 
\hline
\rowcolor[rgb]{0.8,0.902,0.902} \multicolumn{1}{l|}{\textbf{Total}}                                         &                                   & \textbf{337}                                                                                      & \textbf{276 (81.9)}              & \textbf{19 (5.6)}                 & \textbf{42 (12.5)}  & \textbf{250 (74.2)}              & \textbf{73 (21.7)}                & \textbf{14 (4.2)} & \textbf{229 (80.9)}                  \\ 
\hline
\multirow{6}{*}{\rotatebox{90}{\begin{tabular}[c]{@{}c@{}}\textsc{RustRepo}\\ \textsc{Trans}\end{tabular}}} & charset                           & 33                                                                                                & 20 (60.6)                        & 13 (39.4)                         & 0 (0)               & 11 (33.3)                        & 21 (63.6)                         & 1 (3)             & 20 (62.5)                            \\
                                                                                                            & deltachat                         & 125                                                                                               & 54 (43.2)                        & 69 (55.2)                         & 2 (1.6)             & 30 (24)                          & 91 (72.8)                         & 4 (3.2)           & 87 (73.1)                            \\
                                                                                                            & iceberg-java                      & 25                                                                                                & 9 (36)                           & 16 (64)                           & 0 (0)               & 4 (16)                           & 20 (80)                           & 1 (4)             & 20 (83.3)                            \\
                                                                                                            & iceberg-py                        & 44                                                                                                & 15 (34.1)                        & 28 (63.6)                         & 1 (2.3)             & 6 (13.6)                         & 36 (81.8)                         & 2 (4.5)           & 34 (82.9)                            \\
                                                                                                            & crypto-c                          & 20                                                                                                & 16 (80)                          & 4 (20)                            & 0 (0)               & 10 (50)                          & 7 (35)                            & 3 (15)            & 14 (82.4)                            \\
                                                                                                            & crypto-java                       & 97                                                                                                & 39 (40.2)                        & 58 (59.8)                         & 0 (0)               & 26 (26.8)                        & 68 (70.1)                         & 3 (3.1)           & 82 (87.2)                            \\ 
\hline
\rowcolor[rgb]{0.8,0.902,0.902} \multicolumn{1}{l|}{\textbf{Total}}                                         &                                   & \textbf{344}                                                                                      & \textbf{153 (44.5)}              & \textbf{188 (54.7)}               & \textbf{3 (0.9)}    & \textbf{87 (25.3)}               & \textbf{243 (70.6)}               & \textbf{14 (4.1)} & \textbf{257 (78.6)}                  \\ 
\hline\hline
\rowcolor[rgb]{0.902,0.902,0.902} \multicolumn{1}{l|}{\textbf{Total}}                                       &                                   & \textbf{2219}                                                                                     & \textbf{1140 (51.4)}             & \textbf{449 (20.2)}               & \textbf{630 (28.4)} & \textbf{1452 (65.4)}             & \textbf{708 (31.9)}               & \textbf{59 (2.7)} & \textbf{1109 (71.6)}                 \\
\hline
\end{tabular}
    \label{table:rq1-validation-qwen}
\end{table}

\section{Cost Analysis}
\label{sec:cost-analysis}

Table~\ref{table:cost-analysis} shows the cost analysis of different components in \approach for two models, namely, \texttt{\small Claude 3.7 Sonnet} and \texttt{\small Qwen3-Next-80B-A3B}. The results indicate the open-source Qwen3 model is on average $\times23.8$ cheaper than Claude. Please note that we rely on the numbers reported by Claude Code for input and output tokens, and calculate dollar (\$) cost accordingly. For instance Claude requires $\$3$ / $1$ million input tokens and $\$15$ / $1$ million output tokens, and Qwen3 requires $\$0.09$ / $1$ million input tokens and $\$0.78$ / $1$ million output tokens. Since we run our Qwen3 experiments using a proxy server (e.g., LiteLLM~\cite{litellm}) for compatibility purposes, the number of cached input tokens were not reported properly and as a result there is a significant incorrect difference, for instance, ~$2M$ compared to $166$ average input tokens for Test Generator and Repair agent across all tools. However, after checking the logs from the LLM provider, we noticed that the cost was correctly reflected, in the same order of magnitude, for the open-source Qwen3 model.

\begin{table}[!htbp]
    \scriptsize
    \centering
    \caption{Cost analysis of \approach components. Tuple entries indicate $\langle$\texttt{\small Claude 3.7 Sonnet}, \texttt{\small Qwen3-Next-80B-A3B}$\rangle$.}
    \resizebox{\textwidth}{!}{
        \begin{tabular}{c|c|c|c|c|c|c} 
\hline
\textbf{Tool}                   & \begin{tabular}[c]{@{}c@{}}\textbf{Total \#}\\\textbf{Trans. Pairs}\end{tabular} & \textbf{Component} & \textbf{Avg Input Tokens}              & \textbf{Avg Output Tokens}            & \textbf{Avg Duration (ms)}               & \textbf{Avg Cost (\$)}                  \\ 
\hline
\multirow{8}{*}{\alphatrans}    & \multirow{8}{*}{1346}                                                            & Control Flow             & $\langle$437.0, 349.5$\rangle$     & $\langle$86.9, 65.4$\rangle$      & $\langle$2557.5, 2557.5$\rangle$     & $\langle$0.00261, 0.00008$\rangle$  \\
                                &                                                                                  & Data Flow                & $\langle$1124.5, 931.8$\rangle$    & $\langle$122.8, 93.9$\rangle$     & $\langle$3507.8, 3507.8$\rangle$     & $\langle$0.00522, 0.00016$\rangle$  \\
                                &                                                                                  & IO                       & $\langle$782.4, 646.0$\rangle$     & $\langle$205.0, 217.5$\rangle$    & $\langle$7059.7, 7059.7$\rangle$     & $\langle$0.00542, 0.00023$\rangle$  \\
                                &                                                                                  & Lib Equivalence          & $\langle$700.4, 583.0$\rangle$     & $\langle$235.9, 218.2$\rangle$    & $\langle$6900.2, 6900.2$\rangle$     & $\langle$0.00564, 0.00022$\rangle$  \\
                                &                                                                                  & Exception Error          & $\langle$726.4, 599.0$\rangle$     & $\langle$152.0, 182.2$\rangle$    & $\langle$6250.3, 6250.3$\rangle$     & $\langle$0.00446, 0.00020$\rangle$  \\
                                &                                                                                  & Spec                     & $\langle$726.4, 599.0$\rangle$     & $\langle$311.1, 318.6$\rangle$    & $\langle$8801.8, 8801.8$\rangle$     & $\langle$0.00685, 0.00030$\rangle$  \\
                                &                                                                                  & Test Gen \& Repair                 & $\langle$161.6, 2104327.0$\rangle$ & $\langle$11527.8, 7672.8$\rangle$ & $\langle$274732.6, 294980.7$\rangle$ & $\langle$0.17340, 0.19537$\rangle$  \\
                                &                                                                                  & Verdict                  & $\langle$36.6, 32582.1$\rangle$    & $\langle$1404.6, 240.2$\rangle$   & $\langle$31737.7, 9770.2$\rangle$    & $\langle$0.02118, 0.00312$\rangle$  \\ 
\hline
\multirow{8}{*}{\oxidizer}      & \multirow{8}{*}{192}                                                             & Control Flow             & $\langle$1426.4, 1186.9$\rangle$   & $\langle$230.8, 168.5$\rangle$    & $\langle$3957.4, 3957.4$\rangle$     & $\langle$0.00774, 0.00024$\rangle$  \\
                                &                                                                                  & Data Flow                & $\langle$2298.7, 1928.8$\rangle$   & $\langle$206.6, 159.1$\rangle$    & $\langle$3726.5, 3726.5$\rangle$     & $\langle$0.00999, 0.00030$\rangle$  \\
                                &                                                                                  & IO                       & $\langle$993.8, 833.6$\rangle$     & $\langle$238.7, 297.3$\rangle$    & $\langle$6532.9, 6532.9$\rangle$     & $\langle$0.00656, 0.00031$\rangle$  \\
                                &                                                                                  & Lib Equivalence          & $\langle$912.8, 770.6$\rangle$     & $\langle$265.7, 250.3$\rangle$    & $\langle$5640.9, 5640.9$\rangle$     & $\langle$0.00672, 0.00026$\rangle$  \\
                                &                                                                                  & Exception Error          & $\langle$937.8, 786.6$\rangle$     & $\langle$248.3, 240.8$\rangle$    & $\langle$6100.1, 6100.1$\rangle$     & $\langle$0.00654, 0.00026$\rangle$  \\
                                &                                                                                  & Spec                     & $\langle$937.8, 786.6$\rangle$     & $\langle$399.7, 408.7$\rangle$    & $\langle$10097.8, 10097.8$\rangle$   & $\langle$0.00881, 0.00039$\rangle$  \\
                                &                                                                                  & Test Gen \& Repair                 & $\langle$165.2, 2415982.5$\rangle$ & $\langle$11340.1, 8769.0$\rangle$ & $\langle$244057.2, 281051.7$\rangle$ & $\langle$0.17060, 0.22428$\rangle$  \\
                                &                                                                                  & Verdict                  & $\langle$43.1, 28828.1$\rangle$    & $\langle$1813.4, 264.8$\rangle$   & $\langle$37238.3, 6736.6$\rangle$    & $\langle$0.02733, 0.00280$\rangle$  \\ 
\hline
\multirow{8}{*}{\rustrepotrans} & \multirow{8}{*}{344}                                                             & Control Flow             & $\langle$2247.7, 1757.5$\rangle$   & $\langle$266.4, 187.0$\rangle$    & $\langle$5437.8, 5437.8$\rangle$     & $\langle$0.01074, 0.00030$\rangle$  \\
                                &                                                                                  & Data Flow                & $\langle$5083.8, 4841.3$\rangle$   & $\langle$275.7, 186.1$\rangle$    & $\langle$4229.3, 4229.3$\rangle$     & $\langle$0.01939, 0.00058$\rangle$  \\
                                &                                                                                  & IO                       & $\langle$1167.0, 939.1$\rangle$    & $\langle$279.0, 307.9$\rangle$    & $\langle$7104.9, 7104.9$\rangle$     & $\langle$0.00769, 0.00032$\rangle$  \\
                                &                                                                                  & Lib Equivalence          & $\langle$1086.0, 876.1$\rangle$    & $\langle$318.4, 275.4$\rangle$    & $\langle$6532.2, 6532.2$\rangle$     & $\langle$0.00803, 0.00029$\rangle$  \\
                                &                                                                                  & Exception Error          & $\langle$1111.0, 892.1$\rangle$    & $\langle$287.8, 253.5$\rangle$    & $\langle$6326.0, 6326.0$\rangle$     & $\langle$0.00765, 0.00028$\rangle$  \\
                                &                                                                                  & Spec                     & $\langle$1111.0, 889.8$\rangle$    & $\langle$491.5, 457.2$\rangle$    & $\langle$9488.8, 9488.8$\rangle$     & $\langle$0.01071, 0.00044$\rangle$  \\
                                &                                                                                  & Test Gen \& Repair                 & $\langle$189.7, 1329259.2$\rangle$ & $\langle$15020.1, 4355.7$\rangle$ & $\langle$401964.1, 148608.3$\rangle$ & $\langle$0.22587, 0.12303$\rangle$  \\
                                &                                                                                  & Verdict                  & $\langle$44.3, 84397.6$\rangle$    & $\langle$1925.5, 327.8$\rangle$   & $\langle$44992.7, 10575.6$\rangle$   & $\langle$0.02902, 0.00785$\rangle$  \\ 
\hline
\multirow{8}{*}{\skel}          & \multirow{8}{*}{337}                                                             & Control Flow             & $\langle$1163.1, 949.7$\rangle$    & $\langle$201.4, 153.5$\rangle$    & $\langle$3640.0, 3640.0$\rangle$     & $\langle$0.00651, 0.00021$\rangle$  \\
                                &                                                                                  & Data Flow                & $\langle$3621.2, 4951.7$\rangle$   & $\langle$258.6, 195.3$\rangle$    & $\langle$4867.5, 4867.5$\rangle$     & $\langle$0.01474, 0.00060$\rangle$  \\
                                &                                                                                  & IO                       & $\langle$907.7, 761.1$\rangle$     & $\langle$218.5, 245.6$\rangle$    & $\langle$5834.6, 5834.6$\rangle$     & $\langle$0.00600, 0.00026$\rangle$  \\
                                &                                                                                  & Lib Equivalence          & $\langle$826.7, 698.1$\rangle$     & $\langle$233.5, 221.2$\rangle$    & $\langle$5237.1, 5237.1$\rangle$     & $\langle$0.00598, 0.00024$\rangle$  \\
                                &                                                                                  & Exception Error          & $\langle$851.7, 714.1$\rangle$     & $\langle$157.5, 192.7$\rangle$    & $\langle$4896.8, 4896.8$\rangle$     & $\langle$0.00492, 0.00021$\rangle$  \\
                                &                                                                                  & Spec                     & $\langle$851.7, 714.1$\rangle$     & $\langle$368.4, 378.8$\rangle$    & $\langle$8539.1, 8539.1$\rangle$     & $\langle$0.00808, 0.00036$\rangle$  \\
                                &                                                                                  & Test Gen \& Repair                 & $\langle$160.8, 1843194.0$\rangle$ & $\langle$11803.7, 8944.5$\rangle$ & $\langle$245430.8, 272173.9$\rangle$ & $\langle$0.17754, 0.17286$\rangle$  \\
                                &                                                                                  & Verdict                  & $\langle$38.8, 57464.5$\rangle$    & $\langle$1421.1, 355.8$\rangle$   & $\langle$37516.6, 9687.9$\rangle$    & $\langle$0.02143, 0.00545$\rangle$  \\ 
\hline\hline
\multirow{8}{*}{Average}          & \multirow{8}{*}{2219}                                                            & Control Flow             & $\langle$913.6, 731.4$\rangle$     & $\langle$144.6, 106.6$\rangle$    & $\langle$3289.5, 3289.5$\rangle$     & $\langle$0.00491, 0.00015$\rangle$  \\
                                &                                                                                  & Data Flow                & $\langle$2219.1, 2234.6$\rangle$   & $\langle$174.4, 129.2$\rangle$    & $\langle$3845.1, 3845.1$\rangle$     & $\langle$0.00927, 0.00030$\rangle$  \\
                                &                                                                                  & IO                       & $\langle$879.3, 725.1$\rangle$     & $\langle$221.4, 242.7$\rangle$    & $\langle$6835.0, 6835.0$\rangle$     & $\langle$0.00596, 0.00025$\rangle$  \\
                                &                                                                                  & Lib Equivalence          & $\langle$797.7, 662.1$\rangle$     & $\langle$250.9, 230.3$\rangle$    & $\langle$6481.6, 6481.6$\rangle$     & $\langle$0.00616, 0.00024$\rangle$  \\
                                &                                                                                  & Exception Error          & $\langle$823.3, 678.1$\rangle$     & $\langle$182.2, 199.9$\rangle$    & $\langle$6043.5, 6043.5$\rangle$     & $\langle$0.00520, 0.00022$\rangle$  \\
                                &                                                                                  & Spec                     & $\langle$823.3, 677.8$\rangle$     & $\langle$355.4, 357.0$\rangle$    & $\langle$8980.5, 8980.5$\rangle$     & $\langle$0.00780, 0.00034$\rangle$  \\
                                &                                                                                  & Test Gen \& Repair                 & $\langle$166.1, 1971480.1$\rangle$ & $\langle$12094.9, 7446.6$\rangle$ & $\langle$287352.3, 267620.4$\rangle$ & $\langle$0.18192, 0.18324$\rangle$  \\
                                &                                                                                  & Verdict                  & $\langle$38.7, 44068.9$\rangle$    & $\langle$1523.3, 273.4$\rangle$   & $\langle$35146.1, 9620.0$\rangle$    & $\langle$0.02296, 0.00418$\rangle$  \\
\hline
\end{tabular}
    }
    \label{table:cost-analysis}
\end{table}

\section{Optimizing \approach}
\label{sec:optimizing-approach}

There are many directions that can be explored to optimize both the dollar cost and the runtime of \approach, which we outline below, but leave this as future work. However, please note though that \approach’s current dollar cost and runtime compare favorably to the development cost of existing repository-level validation techniques, or even when compared to a salaried human software engineer manually performing validation. Compared to these two baselines, $309$s and $\$1.22$ per translation pair are not daunting. A fully-automated translation tool that runs for a week and costs a few $\$100$ is still extremely useful in practice if the translation produced is high quality. The following are potential optimizations:

\begin{enumerate}
    \item Multiple instances of \approach can run concurrently on functions that are not dependent on each other. We can determine function dependencies by constructing a call graph, which is a common program analysis technique. This approach would reduce the total runtime and does not require any changes to the design of \approach.

    \item Use faster, cheaper LLMs or skip analysis for "trivial" translation pairs (e.g., simple \texttt{\small getter} / \texttt{\small setter} functions or those using only primitive types). Heuristics could detect these to avoid expensive semantic analysis or use a cheaper LLM.

    \item A single instance of \approach can process multiple related functions at once. Part of the overhead of running a coding agent is when the agent “orients” itself by reading code files to understand the codebase. By processing multiple related functions at once, we eliminate redundant "orientation" phases, reducing both runtime and cost.
\end{enumerate}

\section{Threats to Validity}
\label{sec:threats}

Similar to prior techniques, \approach comes with some limitations and threats to the validity. In this section, we discuss how we mitigated various threats.

\vspace{3pt}
\noindent\textbf{Internal Validity.}
There are two main threats to internal validity. First, we only run experiments once. Since LLMs are inherently non-deterministic, running experiments again may produce different results. While it is highly \textit{likely} some individual equivalence verdicts and repair results would change if experiments were run again, it is highly \textit{unlikely} the aggregate metrics we report would change significantly given the large number of translation samples we use ($2{,}219$ pairs). Second, our human investigation does not assess ground truth equivalence. We only assess whether an inequivalent verdict was correct, but we do not analyze the correctness of equivalent verdicts. While this means we don't have any measure of true accuracy of \approach, we still can claim \approach is more accurate than existing automated validation techniques.

\vspace{3pt}
\noindent\textbf{External Validity.}
One main external threat is the generalizability of our approach. Our validation and repair system is very generic and can be extended to more PL pairs with minimal engineering effort. Also, the majority of tools that we used, for example, Tree-Sitter~\cite{tree-sitter} can support a large set of PLs. To mitigate external validity, we built the initial version of \approach with six PLs.

\vspace{3pt}
\noindent\textbf{Construct Validity.}
In order to minimize construct validity, \approach is built on well-known and rigorously tested tools, e.g., Tree-Sitter~\cite{tree-sitter}, Claude Code~\cite{claudecode}, and Codex~\cite{codex}.

\section{Figures}
\label{sec:appendix-figures}

\begin{figure}[!htbp]
    \centering

    \begin{minipage}[t]{0.325\textwidth}
        \begin{monotextbox}
        \tiny
        \input{Resources/Figures/structure-1}
        \end{monotextbox}
    \end{minipage}
    \hfill
    \begin{minipage}[t]{0.325\textwidth}
        \begin{monotextbox}
        \tiny
        \input{Resources/Figures/structure-2}
        \end{monotextbox}
    \end{minipage}
    \hfill
    \begin{minipage}[t]{0.325\textwidth}
        \begin{monotextbox}
        \tiny
        \input{Resources/Figures/structure-3}
        \end{monotextbox}
    \end{minipage}

    \caption{CFG and DFP structures extracted by the Semantic Analyzer component in \approach.}
    \label{fig:cfg-dfg-structure}
\end{figure}

\begin{figure}[!htbp]
    \centering
    \begin{monotextbox}
    \tiny
    \input{Resources/Figures/prompt-template}
    \end{monotextbox}
    \caption{Prompt structure of the Test Generator and Repair Agent.}
    \label{fig:test-gen-prompt-template}
\end{figure}


\end{document}